\documentclass[11pt]{article}
\usepackage[utf8]{inputenc}
\usepackage[top=1in,bottom=1in,left=1in,right=1in]{geometry}
\usepackage{amsmath}
\usepackage{physics}
\usepackage{
authblk}
\usepackage[english]{babel}
\usepackage{amsthm}
\usepackage{amssymb}
\usepackage{hyperref}
\usepackage[capitalise]{cleveref}
\usepackage{tikz}
\usepackage{comment}
\usepackage{booktabs}
\usepackage{tikz}
\usetikzlibrary{quantikz}
\usepackage{glossaries}
\usepackage{lineno}

\newglossary{symbols}{sym}{sbl}{List of Abbreviations and Symbols}
\makeglossary

\newglossaryentry{fn}{type=symbols,name={\ensuremath{F_n}},sort=fn,
description={Empirical (sample) distribution function}}

\newtheorem{theorem}{Theorem}[section]

\theoremstyle{definition}
\newtheorem{definition}{Definition}[section]

\theoremstyle{remark}

\newcommand{\fc}[2]{{#1}^{\dagger}_{#2}}
\newcommand{\fan}[2]{{#1}_{#2}}
\newcommand{\wstat}{\ket{\mathcal{F}}}
\newcommand{\ws}{j}
\newcommand{\sH}{\mathcal{H}}
\newcommand{\oH}{O_{\mathcal{H}}}

\newcommand{\dl}[1]{{{\color{blue}{[DL: #1]}}}}
\newcommand{\dlt}[1]{{{\color{blue}{ #1}}}}
\newcommand{\cy}[1]{{{\color{red}{[CY: #1]}}}}
\newcommand{\wj}[1]{{{\color{magenta}{[WJ: #1]}}}}
\newcommand{\LL}[1]{{{\color{orange}{[LL: #1]}}}}

\newcommand{\del}[1]{{{\color{green}{[suggest to delete: #1]}}}}

\author[1]{Diyi Liu}
\affil[1]{Department of Mathematics, University of Minnesota, Twin Cities, MN}
\author[2]{Weijie Du}
\affil[2]{Department of Physics, Iowa State University of Minnesota, Ames, IA}
\author[3,4]{Lin Lin}
\affil[3]{Department of Mathematics,  University of California, Berkeley, CA}
\author[2]{James P. Vary}
\author[4]{Chao Yang}
\affil[4]{Applied Mathematics and Computational Research Division,  Lawrence Berkeley National Laboratory, Berkeley, CA}

\title{An Efficient Quantum Circuit for Block Encoding a Pairing Hamiltonian}

\begin{document}

\maketitle
\begin{abstract}
We present an efficient quantum circuit for block encoding pairing Hamiltonians often studied in nuclear physics. Our block encoding scheme does not require mapping the creation and annihilation operators to the Pauli operators and representing the Hamiltonian as a linear combination of unitaries. Instead, we show how to encode the Hamiltonian directly using controlled swap operations. We analyze the gate complexity of the block encoding circuit and show that it scales polynomially with respect to the number of qubits required to represent a quantum state associated with the pairing Hamiltonian. We also show how the block encoding circuit can be combined with the quantum singular value transformation to construct an efficient quantum circuit for approximating the density of states of a pairing Hamiltonian. 
The techniques presented can be extended to encode more general second-quantized Hamiltonians.

\end{abstract} \hspace{10pt}

\section{Introduction}
\label{sec:intro}
Block encoding is a recently developed technique for constructing Hamiltonian input model and enabling numerical linear algebra problems to be solved efficiently on a quantum computer~\cite{camps2022explicit,low2017optimal,low2019hamiltonian}.  By embedding a properly scaled matrix $A$ into a larger unitary matrix $U$, one can perform the matrix vector multiplication $Ax$ efficiently on a quantum computer by applying $U$ to a quantum state prepared as $\ket{x}\ket{0}$ provided that $U$ can be efficiently decomposed into simpler unitaries that form the building blocks (quantum gates) of a quantum circuit. When combined with quantum signal processing~\cite{low2017optimal} and quantum eigenvalue/singular value transformation~\cite{gilyen2019quantum}, which allow us to block encode a matrix function of $A$ in terms of the block encoding of $A$, we can solve a variety of numerical linear algebra problems such as linear systems of equations, least squares problems and eigenvalue problems on a quantum computer~\cite{nielsen2001quantum,lin2022lecture,childs2017lecture}.

A general block encoding scheme for sparse matrices was proposed in~\cite{gilyen2019quantum}. The construction of explicit block encoding quantum circuits has been shown in~\cite{camps2022explicit} for a few specific types of sparse matrices. Such constructions have also been developed for other types of sparse matrices in~\cite{sunderhauf2023block,WanBlock,loke2017efficient}. The explicit circuit construction for block encoding can be compressed for noisy intermediate-scale quantum (NISQ) devices~\cite{camps2022algebraic,camps2022fable,camps2020approximate}. A hardware-efficient  technique named Hamiltonian embedding has also been proposed for NISQ quantum computing~\cite{leng2024expanding}. Meanwhile, even if a matrix or an operator is dense, it can still be efficiently block encoded provided there exists special structure, for example, the matrix representation under another basis set is sparse~\cite{zhou2017efficient,li2023efficient}. 

There is considerable interest in constructing efficient block encoding circuits for a second-quantized many-body Hamiltonian~\cite{wan2021exponentially,babbush2018encoding}
\begin{equation}
  \mathcal{H}=\sum_{i,j} h_{ij} \fc{c}{i}\fan{c}{j} + \sum_{i<j,\  k<l} h_{ijkl} \fc{c}{i}\fc{c}{j} \fan{c}{l} \fan{c}{k},
  \label{eq:h2quant}
\end{equation}
that arises in quantum chemistry and nuclear physics,
where $h_{ij} $ and $h_{ijkl}$ denote the strength parameters that correspond to the one-body and two-body interactions of the electrons or nucleons. $c^{\dagger}_i$ and $c_i$ represent the fermionic creation and annihilation operators, with the index $i$ labeling the single-fermion state. The Hamiltonian block-encoding scheme can be utilized in efficient quantum algorithms to perform dynamics simulations and structure calculations ~\cite{du2023multi,babbush2017low,babbush2018low,chan2023simulating}.

%
A widely used technique to block encode $\mathcal{H}$ is to transform the creation and annihilation operators to Pauli operators through the Jordan-Wigner~\cite{nielsen2005fermionic} or Bravyi-Kitaev~\cite{seeley2012bravyi,tranter2018comparison} transformation, which enables one to rewrite \eqref{eq:h2quant} as a linear combination of Pauli unitaries.  By combining a properly constructed preparation oracle circuit that encodes of the coefficients for different terms in Eq. \eqref{eq:h2quant} and a selection orcale circuit that encodes the composition of Pauli operators, one can construct a quantum circuit for encoding a second quantized Hamiltonian. 

In this paper, we present an alternative block encoding scheme for the pairing Hamiltonian that does not require mapping the creation and annihilation operators to Pauli operators.
This scheme follows the procedure laid out in~\cite{camps2022explicit} to construct an explicit block encoding circuit of $\sH$ by treating it as a sparse matrix. 
In particular, we show how a unitary oracle circuit, which we denote by $O_C$ later in this paper, that encodes the nonzero structure of $\sH$ can be constructed through the use of multi-qubit controlled swaps. The nonzero matrix elements of $\sH$ can be encoded by another unitary oracle circuit, which we denote by $\oH$ later in the paper. This circuit can be implemented as a sequence of controlled rotations.  

We demonstrate such a block encoding scheme and the corresponding quantum circuit by applying it to a simple pairing Hamiltonian for many-nucleon systems \cite{Jensen2017,Suhonen2010}. The nuclear pairing interaction is a short-range attraction between nucleons. It contributes to the odd-even effect in the nuclear mass spectrum: that the total binding energy of an odd-$A$ nucleus is less than the average of the total binding energies of the two neighboring nuclei with mass numbers $A\pm 1$ \cite{Suhonen2010}. Besides the effects on nuclear binding energy, the nuclear pairing interaction also plays important roles on various other nuclear properties, such as the moment of inertia, the level-statistics of low-lying states, and the ground state distribution of nucleons in nuclei. The nuclear pairing interaction results from the pairwise correlations between nucleons, which is analogous to the correlated electron pairs in a metallic superconductor in condensed matter physics.

The pairing Hamiltonian \cite{Suhonen2010} can be written as
\begin{equation}
    {\sH}_{\rm pair} = \sum _{\alpha } \sum _{\beta } \sum _{m_{\alpha }>0} \sum _{m_{\beta}>0} G(\alpha , m_{\alpha}, \beta , m_{\beta} )  c^{\dagger } _{\alpha,m_{\alpha} } c^{\dagger} _{\alpha,-m_{\alpha}} c_{\beta,-m_{\beta}} c_{{\beta},m_{\beta}} ,
    \label{eq:pairingModel_fermions}
\end{equation}
where $G(\alpha , m_{\alpha}, \beta , m_{\beta} )$ denotes the interaction coefficient. 
The creation operator $c^{\dagger } _{\alpha,m_{\alpha} } $ creates a state labelled as $\ket{\alpha , m_{\alpha }}$ with $\alpha$ denoting a set of quantum numbers (radial $n_\alpha$, orbital angular momentum $l_\alpha$ and total angular momentum $j_\alpha$) associated with a single-particle state and $ m_{\alpha } $ being the absolute value of the projection of $j_{\alpha }$. Similarly, $c^{\dagger} _{\alpha,-m_{\alpha}} $ creates the state labelled by $\ket{\alpha , - m_{\alpha }}$. We refer to the states $\ket{\alpha , m_{\alpha }}$ and $\ket{\alpha , - m_{\alpha }}$ as the paired single-particle states that are created or annihilated jointly. The individual monomial $c^{\dagger } _{\alpha,m_{\alpha} } c^{\dagger} _{\alpha,-m_{\alpha}} c_{\beta,-m_{\beta}} c_{{\beta},m_{\beta}}$ in $\mathcal{H} _{\rm pair}$ first annihilates the paired single-particle states $\ket{\beta , m_{\beta }}$ and $\ket{\beta , - m_{\beta }}$ and then creates the paired states $\ket{\alpha , m_{\alpha }}$ and $\ket{\alpha , - m_{\alpha }}$ in a correlated manner.

For fermions, the creation and annihilation operators in \eqref{eq:pairingModel_fermions} are subject to the anticommutation relations
\begin{align}
    \{ c^{\dagger } _{\alpha, \pm m_{\alpha} }, \ c _{\beta, \pm m_{\beta} } \} = \delta _{\alpha , \beta} \delta _{m_{\alpha}, m_{\beta}}, \ \{ c^{\dagger } _{\alpha, \pm m_{\alpha} }, \  c^{\dagger} _{\beta, \pm m_{\beta}} \} = \{ c_{\alpha, \pm m_{\alpha}}, \  c_{{\beta}, \pm m_{\beta}}  \}  =0. 
\end{align}

To simplify the notations, we define a mapping $\mathcal{Q}$ between the quantum numbers $(n_\alpha,l_\alpha,j_\alpha,m_\alpha)$  associated with a single particle state and a unique integer so that 
\begin{align*}
   \mathcal{Q}(n_\alpha,l_\alpha,j_\alpha,m_\alpha)=2p, \quad  \mathcal{Q}(n_\alpha,l_\alpha,j_\alpha,-m_\alpha)=2p+1 ,
\end{align*}
for a unique integer $p$.  As a result, we can rewrite $c^{\dagger}_{\alpha,m_\alpha}$ simply as $c^{\dagger}_{2p}$ and $c^{\dagger}_{\alpha,-m_\alpha}$ as $c^{\dagger}_{2p+1}$. Similar simplification can be applied to the annihilation operators.  

Since the single particle states indexed by $2p$ and $2p+1$ are created or annihilated together, we can treat the creation or annihilation of these particle pairs as the creation or annihilation of a quasiparticle consisting of a correlated pair of fermions.  We use the notation $a_p^{\dag}$ and $a_p$ to denote the creation and annihilation of such a quasiparticle, respectively. With this notion, we express the pairing Hamiltonian simply as a (quasi) one-body operator
\begin{equation}
\label{eq:Ham1}
    \sH _{\rm pair} = \sum _{p,q} 
    h_{pq}a^{\dagger}_{p} a_{q} .
\end{equation}

This paper is organized as follows. In section \ref{sec:notationsAndConventions}, we introduce the basic notations and conventions.  
We review the basic principles of block encoding and quantum signal processing in section~\ref{sec:basic}.  
In section~\ref{sec:circuit}, we first review the general structure of a block encoding circuit for sparse matrices presented in earlier work~\cite{GrandUni}. We then show how the explicit circuit for the $O_C$ oracle that encodes the nonzero structure of a (pseudo)one-body of Hamiltonian of the form \eqref{eq:pairingModel_fermions} can be constructed by a sequence of multi-qubit swaps. We also show how the circuit for the $O_{\sH}$ oracle that encodes the nonzero matrix elements of $\sH$ can be constructed as a sequence of controlled rotations.  
We give two examples to show how the circuit construction techniques presented in section~\ref{sec:circuit} can be used to block encode a simple toy Hamiltonian and a pairing Hamiltonian in a Fock space defined by a limited number of single-particle basis states.
We also show how such a block encoding circuit for the pairing Hamiltonian can be utilized to construct a quantum circuit that can be used to approximate the density of states of the Hamiltonian for all the systems describable in the chosen basis space. The latter requires us to use quantum signal processing/quantum eigenvalue transformation to contruct a block encoding circuit for a matrix function of $\sH$ in terms of the block encoding circuit of $\sH$.  
Additional comments are presented in section~\ref{sec:conclude}.


\section{Notations and conventions}
\label{sec:notationsAndConventions}
We adopt standard conventions used in the quantum computing literature \cite{nielsen2010quantum} 
and use the Dirac $\bra{\cdot}$ and $\ket{\cdot}$ notation to denote respectively row and column vectors.
In particular
$\ket{0}$ and $\ket{1}$ are used to represent the unit vectors $e_{0} = [1 \ 0]^T$ and $e_{1} = [0 \ 1]^T$, respectively.
The tensor product of $m$ $\ket{0}$'s is denoted by $\ket{0^m}$. We use $\ket{x,y}$ to represent the Kronecker product of $\ket{x}$ and $\ket{y}$, which is also sometimes written as $\ket{x}\ket{y}$ or $\ket{xy}$.
The $N\times N$ identity matrix is denoted by $I_N$ and we sometimes drop the subscript $N$
when the dimension is clear in the context.  

We map a binary representation of an integer $l \in \mathbb{N}: 0 \leq l \leq 2^{n}-1$, 
\begin{equation*}
    l = l_0 \cdot 2^{0} + l_1 \cdot 2^{1} + \dots+ l_{n-1} \cdot 2^{n-1},
\end{equation*}
where $l_i \in \{ 0,1\}$ for $i =0, 1,2 \dots n-1$, to a quantum state $\ket{l}$ often written as $\ket{l_{0}, \dots,l_{n-1}}$. The superposition of these states can be encoded with $n$ qubits where $\ket{l_i}$ is mapped to the $i$th qubit.  

For a many-nucleon system, $\ket{l_{0}, \dots,l_{n-1}}$ can also be interpreted as an occupation representation of a many-body basis in which the value of $l_i$ indicates the occupation of the $i$th single-particle state used to approximate a many-body state within a $2^n$ dimensional Hilbert space (Fock space).

We employ the letters $H$, $X$, $Y$, and $Z$ to represent the Hadamard, Pauli-$X$, Pauli-$Y$, and Pauli-$Z$ matrices, respectively. They are defined as
\begin{align}
H &= \frac{1}{\sqrt{2}} \begin{bmatrix}
1 & \phantom{-}1 \\
1 & -1
\end{bmatrix}, &
X &= \begin{bmatrix}
0 & 1 \\
1 & 0
\end{bmatrix}, &
Y &= \begin{bmatrix}
0 & -i \\
i &  \phantom{-}0
\end{bmatrix}, &
Z &= \begin{bmatrix}
1 &  \phantom{-}0 \\
0 & -1
\end{bmatrix}.
\label{eq:basicgates}
\end{align}
These unitary matrices are used as single qubit gates in quantum computing.

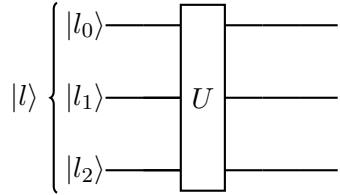
\begin{figure}[htbp]
  \centering
  \begin{quantikz} 
    \lstick[3]{ $\ket{l}$} \ket{l_0} &  \qw & \gate[3]{U} & \qw & \qw & \qw \\
     \ket{l_1} &  \qw & \qw   & \qw & \qw &\qw \\ 
     \ket{l_2} & \qw  & \qw  &\qw & \qw & \qw 
\end{quantikz}
\caption{Illustration of the circuit convention. The circuit is prepared with Quantikz Package~\cite{kay2018tutorial}.}
\label{fig:democirc}
\end{figure}

We follow the standard convention for drawing quantum circuits with multiple parallel lines covered by several layers of rectangular boxes. Each line corresponds to either a single qubit or
multiple qubits depending on how it is labelled and each box corresponds to a single qubit or multi-qubit gate depending on the number of qubit lines passing through it. We use the convention that the qubits in a circuit diagram  are numbered increasingly from the top to the bottom as illustrated by the 3 qubit circuit $U$ in \cref{fig:democirc}. An integer $l = [l_{n-1} \cdots l_1 l_0]$ input to a circuit is prepared as a  quantum state $\ket{l_{0}\ldots l_{n-1}}$ with $\ket{l_0}$ mapped to the $0$th qubit and $\ket{l_{n-1}}$ mapped to the $(n-1)$th qubit.

An important class of quantum gates are the controlled gates, i.e., one or more qubits act as a control for some operation.
Graphically, the control operation is represented by a vertical line connecting the control qubit(s), marked by either a solid or open circle, to the so-called target gate, see e.g., the controlled NOT (CNOT) in \cref{fig:cnotgate}.  The target can also be multiple gates grouped in a subcircuit block used to perform certain operations.
A solid circle indicates that the controlled operation is performed on the connected qubit, when the input to the controlling qubit is a $\ket{1}$ state.
Similarly, an open circle indicates that the controlled operation is performed when the input to the controlling qubit is a $\ket{0}$ state.

\begin{figure}[htbp]
  \centering
  \begin{quantikz} 
  q_{0} & \ctrl{1} & \qw \\
  q_{1} & \targ{} &\qw 
\end{quantikz}
=$\begin{bmatrix}
1 & 0 & 0 & 0 \\
0 & 1 & 0 & 0\\
0 & 0 & 0 & 1\\
0 & 0 & 1 & 0
\end{bmatrix}$ \quad \quad 
 \begin{quantikz} 
  q_{0} & \octrl{1} & \qw \\
  q_{1} & \targ{} &\qw 
\end{quantikz} 
=$\begin{bmatrix}
0 & 1 & 0 & 0 \\
1 & 0 & 0 & 0\\
0 & 0 & 1 & 0\\
0 & 0 & 0 & 1
\end{bmatrix}$
\caption{Illustration of Controlled-Not Gates.}
\label{fig:cnotgate}
\end{figure}

Mathematically, the left and right controlled gates in \cref{fig:cnotgate} denote
\begin{equation}
E_1 \otimes X + (I - E_1) \otimes I,
\qquad \mbox{and} \qquad
E_0 \otimes X + (I - E_0) \otimes I,
\end{equation}
respectively, where the orthogonal projection operators are
\begin{align}
E_{1} &= e_{1} e_{1}^T = \ket{1}\bra{1}, &
E_{0} &= e_{0} e_{0}^T = \ket{0}\bra{0} .
\label{eq:e1e2}
\end{align}
Note that the NOT ($X$) operation is applied to the input of qubit $q_1$ in the left circuit in \cref{fig:cnotgate} only if the input to qubit $q_0$ is $\ket{0}$. Likewise, the NOT operation is applied to the input of qubit $q_1$ in the right circuit in \cref{fig:cnotgate} if the input to qubit $q_0$ is $\ket{1}$. A similar expression can be
use to denote multi-qubit controlled NOT gates.


\section{Block encoding and quantum signal processing}
\label{sec:basic}


Block encoding is a technique for embedding a properly scaled non-unitary 
matrix $A \in \mathbb{C}^{N\times N}$ into a unitary matrix $U_A$ of the form
\begin{equation}
U_A = \begin{bmatrix}
A & \ast \\
\ast & \ast
\end{bmatrix},
\end{equation}
where $\ast$ denotes a matrix block yet to be determined.
Applying $U_A$ to a vector of the form
\begin{equation}
v = \begin{bmatrix}
x \\
0
\end{bmatrix} = \ket{0}\ket{x},
\end{equation}
yields
\[
w = U_A v = \begin{bmatrix}
Ax \\
*
\end{bmatrix} = \ket{0}(A\ket{x}) + \ket{1}\ket{\ast} ,
\]
where $\|x\|_2=1$ and $\ast$ denotes a vector with unimportant information.
If we measure the first qubit and obtain the $\ket{0}$ state, the
second qubit register then contains $A\ket{x}$. The probability of such a 
successful measurement is $\|Ax\|^2$.

A more general definition of block encoding is as follows.
\begin{definition}[Block encoding \cite{camps2022explicit,lin2022lecture}] 
Given an $n$-qubit matrix $A \in \mathbb{C}^{2^n\times 2^n}$, if we find $\alpha, \epsilon \in \mathbb{R}_+$, and an $(m+n)$-qubit unitary matrix $U_{A}$ so that 
\begin{equation}
  || A-\alpha(\bra{0^m} \otimes I_{2^n}) U_{A} (\ket{0^m} \otimes I_{2^n}) ||\leq \epsilon
\end{equation}
then $U_A$ is called an $(\alpha,m,\epsilon)$-block-encoding of $A$. In this paper, we only consider exact block encoding in which $\epsilon=0$ and simply refer to $U_A$ as an $(\alpha,m)$-block-encoding.
\end{definition}

Assuming $U_A$ can be represented by an efficient quantum circuit, we can then apply $A$ to a quantum state $\ket{x}$ on a quantum computer and generate $\ket{Ax}$ through the measurement of the ancilla qubits.

However, to solve the linear algebra problems involving $A$ such as computing its lowest eigenvalue and corresponding eigenvector, we need to block encode a matrix function of $A$ that is often approximated by a matrix polynomial. 
The quantum signal processing theory for scalar polynomials and its extension to matrix polynomials through the quantum singular value or eigenvalue tranformation theory enables one to access the block encoding of a matrix polynomial of $A$ on the basis of the block encoding of $A$.

For completeness, we present the quantum signal processing theory below,
which is a slight variation of \cite[Theorem 4]{gilyen2019quantum}.

\begin{theorem}[Quantum signal processing] \label{thm:qsp}
Let 
\begin{equation}
 U(t) = 
 \begin{bmatrix}
 t & \sqrt{1-t^2} \\
 \sqrt{1-t^2} & -t
 \end{bmatrix},
\label{eq:Ut}
\end{equation}
where $|t|\leq 1$.
There exists a set of phase angles $\Phi_d \equiv \{\phi_0,,...,\phi_d\} \in \mathbb{R}^{d+1}$so that 
\begin{equation}
U_{\Phi_d}(t) \equiv (-i)^{d} e^{i\phi_0 Z} \Pi_{j=1}^d \left[U(t)e^{i\phi_j Z} \right] 
 =
 \begin{bmatrix}
 p(t) & -q(t)\sqrt{1-t^2}\\
 q^*(t)\sqrt{1-t^2} & p^*(t)
 \end{bmatrix},
 \label{eq:Uphi}
\end{equation}
if and only if $p(t)$ and $q(t)$ are complex valued polynomials in $t$ and satisfy
\begin{enumerate}
\item $\mathrm{deg}(p) \leq d$, $\mathrm{deg}(q) \leq d-1$;
\item $p$ has parity $d$ mod $2$ , and $q$ has parity  $d-1$ mod $2$; 
\item $|p(t)|^2 + (1-t^2)|q(t)|^2 = 1$, $\forall t\in [-1,1]$.
\end{enumerate}
When $d=0$, $\deg(q)\le-1$ should be interpreted as $q=0$.
\end{theorem}

\begin{figure}[htbp]
\centering
  \begin{quantikz} 
    \ket{0} &\gate{H}&\gate[2]{CR_{\phi_{d}}}& \qw & \gate[2]{CR_{\phi_{d-1}}}& \qw &\qw \dots & \qw &\gate[2]{CR_{\phi_{0}}} & \gate{H} &\qw \\ 
     \ket{0^{m}} & \qw &\qw &\gate[2]{U_{A}} & \qw & \gate[2]{U_{A}^{\dagger}} & \qw \dots & \gate[2]{U_{A}} & \qw & \qw &\qw \\
    \ket{\psi}  &  \qw & \qw & \qw &\qw & \qw &\dots & \qw &\qw &\qw \\
\end{quantikz}
\caption{Illustration: the schematic circuit design of quantum singular value transformation (for an odd $d$; for an even $d$ the last $U_A$ is replaced by $U_A^{\dag}$). The additional Hadamard gate selects only the real part of the polynomial $p$. 
}
\label{fig:QSP}
\end{figure}
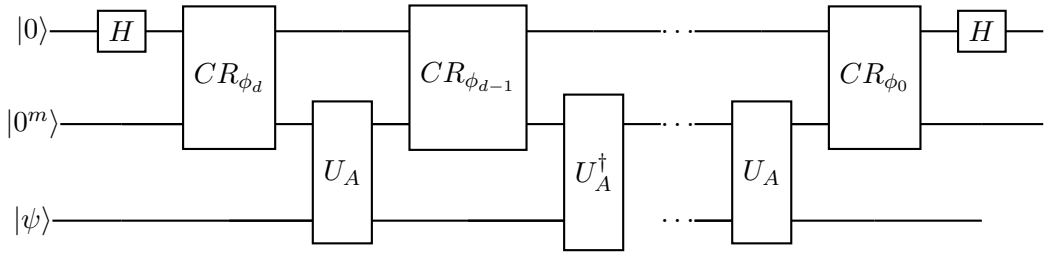

Note that $U_{\Phi_d}(t)$ is a block encoding of the function $p$ of a scalar $t$ in Theorem \ref{thm:qsp}. The theorem can be  extended to a properly scaled matrix which is the theory behind quantum singular value transform (QSVT). 
When the block encoding of $A$, denoted by $U_A$,  is Hermitian, QSVT reduces to quantum eigenvalue transformation (QET). It follows from QSVT that the block encoding of $p(A)$ for some polynomial $p(t)$ can be expressed in terms of the block encoding of $A$ as illustrated in \cref{fig:QSP}. In the circuit representation, $\phi_i$ is the same phase angle appearing in the Theorem \ref{thm:qsp}. Each controlled-rotation gate $CR_{\phi_{i}}$ can be implemented as a single-qubit gate placed on the top ancilla qubit that performs $e^{-i\phi_{i}Z}$, preceeded and followed by a multi-qubit CNOT gate~\cite{lin2022lecture}. We show the QSVT circuit for a specific example in the section \ref{sec:computedos}. Although Theorem \ref{thm:qsp} only states the existence of such phase angles for quantum signal processing, these phase angles can be efficiently computed by solving a numerical optimization problem with a specific symmetric initial condition for phase angles \cite{dong2022infinite,Dongefficient}. 

%
%

\section{Block encoding circuit}
\label{sec:circuit}
In this section, we describe how the second-quantized pairing Hamiltonian \eqref{eq:Ham1} can be block encoded and how an efficient quantum circuit can be constructed. When we consider a general pairing Hamiltonian including self-energy term $\fc{c}{p} \fan{c}{p}$, it is necessary to directly block encode the Hamiltonian written with creation and annihilation operator of fermions \eqref{eq:pairingModel_fermions} (see section \ref{sec:examples} for more details).
To keep an concise notation for the creation and annilation operator, we rewrite the Hamiltonian \ref{eq:Ham1} and use an integer pair to label multiplication of a creation and an annihilation as shown in \cref{eq:Ham1},
\begin{equation}
\label{eq:hpairnumber}
    \mathcal{H} _{\rm pair} = \sum _{p,q} 
    h_{pq}a^{\dagger}_{p} a_{q} \equiv \sum_{l}v_{l}\mathcal{H}_{l}.
\end{equation}
with the pair index $l \in \{ (p,q)| 0 \leq p \leq n-1, 0\leq q \leq n-1 \}$. For each $l=(p,q)$, $v_{l}$ is defined as $h_{pq}$ in \eqref{eq:hpairnumber} and $\mathcal{H}_{l}$ is defined as $a^{\dagger} _{p } a_{q}$. The relabeling does not lose any generality as a one-to-one mapping can be defined for the two sets notation for labeling fermionic system.
We view $ \mathcal{H} _{\rm pair} $ as a sparse matrix.
A general recipe for constructing a block encoding of an $s$-sparse matrix, which is defined to be a matrix that has at most $s$ nonzero matrix elements in each column, is described by the following theorem (see~\cite{camps2022explicit,gilyen2019quantum}).

\begin{theorem} \label{thm:Usparse}
Let $c(j,\ell)$ be a function that gives the row index of the $\ell$th (among a list of $s$) non-zero matrix elements in the $j$th column of an $s$-sparse matrix $A \in \mathbb{C}^{N\times N}$ with $N=2^n$, where $s=2^m$.  If there exists a unitary $O_c$ such that 
\begin{equation}
O_c \ket{\ell}\ket{j} = \ket{\ell}\ket{c(j,\ell)},
\label{eq:oc}
\end{equation}
and a unitary $O_A$ such that
\begin{equation}
O_A \ket{0}\ket{\ell}\ket{j} = \left(
A_{c(j,\ell),j} \ket{0} + \sqrt{1-|A_{c(j,\ell),j}|^2} \ket{1}
\right)
\ket{\ell}\ket{j},
\label{eq:oa}
\end{equation}
then 
\begin{equation}
U_A = \left(I_2 \otimes D_s\otimes I_N \right) 
\left(I_2 \otimes O_c \right) 
O_A
\left(I_2 \otimes D_s \otimes I_N\right),
\label{eq:uafact}
\end{equation}
block encodes $A/s$.
Here $D_s$ is called a diffusion operator and is defined as 
\begin{equation}
D_s \equiv \underbrace{H\otimes H \otimes \cdots \otimes H}_{m},
\end{equation}
\end{theorem}

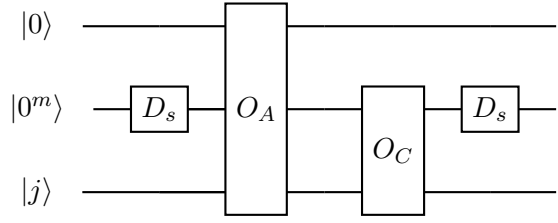
\begin{figure}
  \begin{center}
    \begin{quantikz}
    &\ket{0} \quad & \qw            & \gate[3]{O_{A}} &\qw  &\qw &\qw &\qw \\ 
    &\ket{0^m} \quad & \gate{D_s}     & \qw         &\qw &\gate[2]{O_C} & \gate{D_{s}} &\qw\\
    &\ket\ws \quad & \qw            & \qw         &\qw & & \qw &\qw
  \end{quantikz}
  \end{center}
  \caption{Illustration of the circuit for matrix block encoding.}
\label{fig:complete_circ}
\end{figure}

The basic structure of the block encoding circuit is shown in \cref{fig:complete_circ}.

Note that the input to the circuit consists of three set of qubits. The top qubit is used to encode nonzero matrix elements.  The second set of qubits that take $\ket{0^m}$ as the input are used to encode the position of the nonzero elements. The bottom set of qubits that take $\ket{j}$ as the input are the qubits that encodes the column indices of the matrix to be block encoded.

The circuit consists of several blocks. The $O_C$ block is used to encode the positions of the nonzero elements which is defined by the mapping $c(j,\ell)$ in Theorem \ref{thm:Usparse}. The $O_A$ block is used to encode the nonzero matrix elements. 

The structure of $O_C$ depends on the mapping $c(j,\ell)$, which describes the sparsity structure of $A$. Ref. \cite{camps2022explicit} showed how $c(j,\ell)$ is defined for tridiagonal Toeplitz matrices and sparse matrices whose sparsity structure can be described by a binary tree, as well as how the $O_C$ circuit can be constructed efficiently for those type of matrices. 

Once $O_C$ is specified, we can construct $O_A$ which encodes the numerical value of each non-zero matrix elements.




\subsection{The $O_C$ circuit}
In this section, we describe how the $O_C$ circuit can be constructed for the pairing Hamiltonian $ \mathcal{H} _{\rm pair} $ \eqref{eq:hpairnumber}.

\subsubsection{Define $c(j,\ell)$ by bit swap}
\label{sec:bitswap}
The nonzero structure of the Hamiltonian is determined by the composition of the creation and annihilation operators in the second quantization representation of the Hamiltonian \eqref{eq:pairingModel_fermions}. The application of $a^{\dagger} _{p } a_{q}$ to the many-body basis $\ket{j}$ (a single Slater determinant) can potentially create a nonzero matrix element in the $j$th column of $ \mathcal{H} _{\rm pair} $ in the many-body basis.
To understand such a non-zero structure, let us first consider the outcome of applying $a_p^{\dagger}a_q$
to a particular column $\ket{j}$ with a bit representation $\ket{j_0\cdots j_{n-1}}$ where $j_k \in \{0,1\}$.  If $\ket{j}$ is a $m$ particle state, then $m$ of the $n$ qubits are set to $\ket{1}$ while others are in $\ket{0}$. As a result, we can represent the state by an $m$-tuple $(f_0,f_1 \dots, f_{m-1})$ with $f_0$, $f_1, \cdots f_{m-1}$ being the position of qubits in the $\ket{1}$ state. 
Alternatively, we can also represent $\ket{j}$ by
\begin{equation}
  \ket{j} = \fc{a}{f_0} \fc{a}{f_{1}} \dots \fc{a}{f_{m-1}} \ket{vac},
  \label{eq:binaryRepresentationOfFockState}
\end{equation}
with $\ket{vac}$ denoting the vacuum state in which all the basis states are vacant.

If we use $\ket{j}_{p}$ to represent the $p$th bit in the binary representation of $\ket{j}$,  it is easy to verify that for $ \mathcal{H} _{l} = \fc{a}{p}\fan{a}{q}$ defined in \eqref{eq:hpairnumber},
\begin{equation}
\fc{a}{p}\fan{a}{q} \ket{j} = 0,
\end{equation}
unless
\begin{align}
    \begin{cases}
       \ket{j} _q = \ket{1}  \ \text{and} \ \ket{j}_p = \ket{0} \ & \text{for} \ \ \ p \neq q, \\
      \ket{j}_q = \ket{1} \ & \text{for} \ \ \  p = q . 
    \end{cases}
    \label{eq:pqcond}
\end{align}


In fact, if $\ket{j}$ satisfies \eqref{eq:pqcond}, it is easy to show that 
    \begin{align}
    \begin{cases}
       [\fc{a}{p}\fan{a}{q} \ket{j}]_{p}=\ket{1}, \ [\fc{a}{p}\fan{a}{q} \ket{j}]_{q}=\ket{0} \ & \text{for} \ \ \ p \neq q, \\
      [\fc{a}{p}\fan{a}{p} \ket{j}]_{p}=\ket{1} \ & \text{for} \ \ \  p = q . 
    \end{cases}
    \label{eq:bitswap}
\end{align}

Therefore, the operator $\fc{a}{p}\fan{a}{q}$ simply swaps the $p$th and $q$th bits in $\ket{j}$ \eqref{eq:binaryRepresentationOfFockState} (the case with $p=q$ is trivial), creating a non-zero matrix element in the row whose index can be represented by the permuted bit string. 


If $\ket{j}$ satisfies \eqref{eq:pqcond}, the conditioned swap operation associated with $ \mathcal{H} _{\ell}$ defines the $c(j,\ell)$ function in Theorem~\ref{thm:Usparse} as 
\begin{equation}
  c(j,\ell) = \begin{cases}
    \text{SWAP}(\ket{j}, p, q) &\text{if \eqref{eq:pqcond} holds,}\\
    \text{invalid}  &\text{\rm{otherwise}} ,
  \end{cases}
  \label{eq:clj}
\end{equation}
where ${\text{SWAP}}(\ket{j}, p, q)$ denotes the many-body state obtained from the binary representation of $\ket{j}$ by swapping the $p$-th qubit with the $q$-th qubit. When $c(j,\ell)$ is invalid, we can set it to any arbitrary state, e.g., $\ket{j}$, and record the error information with an additional ancilla qubit (see discussions in section~\ref{sec:circuit_for_U}).

We should note that, in general, applying a one-body operator $\fc{a}{p}\fan{a}{q}$ to $\ket{j}$ introduces a phase factor $(-1)^{j_{q+1}+j_{q+2}\dots + j_{p-1}}$. 
Such a phase factor does not appear in encoding each $\mathcal{H}_{\ell}$ in $\mathcal{H}_{\rm pair}$ \eqref{eq:hpairnumber} as $\fc{a}{p}$ and $\fan{a}{q}$ are pseudo one-body operators that represent the creation and annihilation of the paired fermions in practice. In other words, the phase factor is always one as $j_{q+1}+j_{q+2}\dots + j_{p-1}$ is always an even integer in our example. The absence of the phase factors simplifies the $O_C$ circuit construction. 

\subsubsection{The general structure of the $O_C$ circuit}

We note that the mapping between $\ket{j}$ and the valid $\ket{c(j,\ell)}$ involves a conditional swap operation, which is unitary. Therefore, we can use a controlled swap gate to implement $O_C$ for $\mathcal{H}_{\ell}$ associated with a fixed $\ell$. We discuss how such a controlled swap operation, which depends on both $\ket{\ell}$ and $\ket{j}$,  can be implemented in the next section. The implementation requires two additional ancilla qubits to activate the control, and to discard invalid operations according to the possible error information.  

To construct the $O_C$ circuit for $\sum_{\ell=0}^{L-1} \mathcal{H}_{\ell}$, which defines the nonzero structure of $\mathcal{H}$ defined in \eqref{eq:hpairnumber}, we use a  technique often referred to as a \textit{select} oracle defined as
\begin{equation}
\mathrm{select}(\mathcal{H}) = \sum_{\ell=0}^{L-1} \ket{\ell}\bra{\ell} \otimes \mathcal{H}_{\ell},
\end{equation}
where ${\ell}$ corresponds to the pairwise indices $(p,q)$ of the single-particle basis. We encode $\ket{\ell}$ with a set of ancilla qubits that is referred to as the \textit{selection qubits}.
The input to these qubits is the $\ket{0}^m$ state, where $m = \log L$. We use a diffusion operator (a set of Hadamard gates) to create a superposition of all possible $\ell$'s, which enumerate all possible $(p,q)$ pairs.

The selection oracle can be implemented as a product of controlled operations $S_{L-1}\cdots S_1\cdot S_0$, where 
\begin{equation}
S_{\ell} = \ket{\ell}\bra{\ell} \otimes \mathcal{H}_{\ell} +  \left(I-\ket{\ell}\bra{\ell}\right) \otimes I.
\label{eq:controlU}
\end{equation}
Here the projector $\ket{\ell}\bra{\ell}$ serves as the control.
When applied to $\frac{1}{2^{m/2}}\sum_{\ell=0}^{L-1} \ket{\ell}\otimes \ket{j}$, we obtain
\begin{equation}
    \frac{1}{2^{m/2}} S_{L-1}\cdots S_1\cdot S_0 \sum_{\ell=0}^{L-1} \ket{\ell}\otimes \ket{j} =
\frac{1}{2^{m/2}} \sum_{\ell=0}^{L-1}  \ket{\ell}\otimes \mathcal{H}_{\ell} \ket{j}.
\end{equation}

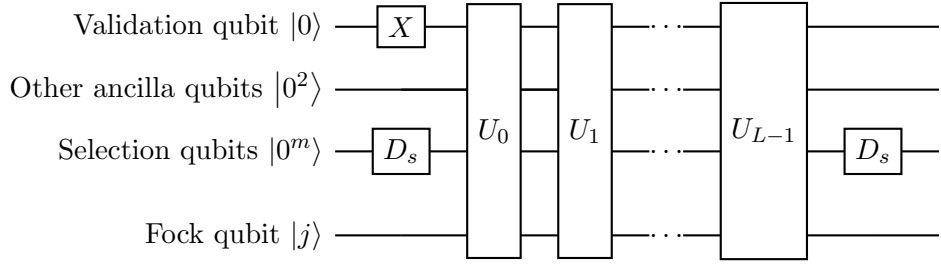
\begin{figure}[htbp]
\centering
  \begin{quantikz} 
    \lstick[1]{Validation qubit $\ket{0}$} &\gate{X}&\gate[4]{U_{0}}&\gate[4]{U_{1}}& \qw \dots & \gate[4]{U_{L-1}} & \qw & \qw\\ 
     \lstick[1]{Other ancilla qubits $\ket{0^2}$} & \qw &\qw &\qw & \qw \dots &  & \qw & \qw \\
    \lstick[1]{Selection qubits $\ket{0^{m}}$}  &  \gate{D_s} & & & \qw \dots &  & \gate{D_s} & \qw \\
    \lstick[1]{Fock qubit $\ket{j}$}  &  \qw & & &\qw \dots &  &\qw & \qw\\
\end{quantikz}
\caption{The basic structure of the $O_C$ circuit. Qubits labeled by ``Other ancilla qubits" consist of a controlling qubit and a rotation qubit as discussed in sections \ref{sec:circuit_for_U} and \ref{sec:Oh}.   }

\label{fig:OCsequence}
\end{figure}

 We combine the control (i.e., $\ket{\ell}\bra{\ell}$) in \eqref{eq:controlU} with the control used to implement \eqref{eq:clj}. In particular, we need to include an ancilla qubit to invalidate operations in which the condition \eqref{eq:pqcond} does not hold, and some additional ancilla qubits to implement the controlled swap. For each $\ell$, these controlled operations collectively define a unitary operator $U_{\ell}$, that can be applied in sequence as shown in the circuit structure presented in Figure~\ref{fig:OCsequence}. Note that in this circuit, the validation qubit is set to $\ket{1}$ by the NOT ($X$) gate in advance.
We assume the diffusion operator has already been applied to the $\ket{0}^{m}$ input to create a superposition of the $\ket{\ell}$ basis in the selection qubits.

\subsubsection{The circuit for $U_{\ell}$}
\label{sec:circuit_for_U}
We now discuss how the circuit for each $U_{\ell}$ can be constructed. As the swap operation is only performed when \eqref{eq:pqcond} is satisfied, such an operation should be carried out as a controlled swap.

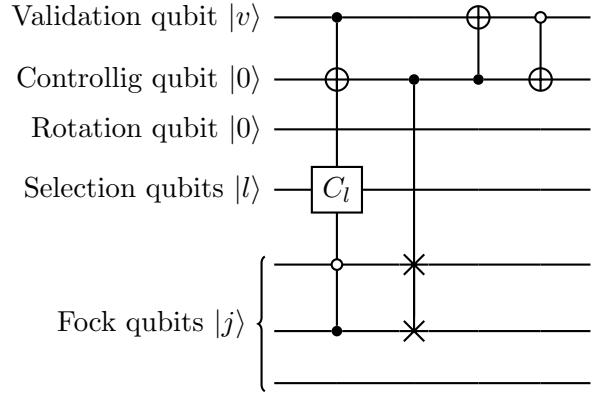
\begin{figure}[htbp]
  \centering
  \begin{quantikz} 
     \lstick[1]{Validation qubit $\ket{v}$} & \ctrl{4} & \qw & \targ{}  & \octrl{1} & \qw\\ 
     \lstick[1]{Controllig qubit $\ket{0}$} &   \targ{}  & \ctrl{4} & \ctrl{-1}   & \targ{} & \qw \\
          \lstick[1]{Rotation qubit $\ket{0}$} &\qw &\qw &\qw  &\qw & \qw \\
    \lstick[1]{Selection qubits $\ket{l}$}  &  \gate{C_l} & \qw & \qw & \qw  & \qw \\
    \lstick[3]{Fock qubits $\ket{j}$} &  \octrl{1} & \swap{1} & \qw & \qw & \qw \\
     &  \ctrl{} & \swap{}   & \qw & \qw &\qw \\ 
     & \qw  & \qw  &\qw & \qw & \qw 
\end{quantikz}
\caption{Illustration of $U_l$ circuit with $\mathcal{H}_l=\fc{a}{p}\fan{a}{q}$ and $p\neq q$.}
\label{fig:oc2}
\end{figure}

The control required to perform a specific swap operation for a particular pair of qubits is defined by both $\ket{j}$ and $\ket{\ell}$, i.e., the corresponding operator $\fc{a}{p}\fan{a}{q}$. Since we can not place controls on the qubits to be swapped directly, we need an ancilla qubit to serve as a \textit{controlling qubit} that activates the swap operation when that qubit is in the $\ket{1}$ state. In Figure~\ref{fig:oc2}, this controlling qubit is the second qubit from the top.  The controlled swap is performed in the third layer from the end of the circuit. 

The controlling qubit is initialized to the $\ket{0}$ state. It is turned to the $\ket{1}$ state when  \eqref{eq:pqcond} holds for the input $\ket{j}$, which can be implemented as a set of controls placed on both $\ket{l}$ and $\ket{j}$ that are compatible.
After a controlled swap operation is performed, we turn the validation qubit from $\ket{1}$ to $\ket{0}$ by a CNOT gate (with a control placed on the controlling ancilla qubit) to indicate that a valid $\mathcal{H}_l\ket{j}$ has been performed. In addition, we use another CNOT gate (with a control placed on the validation qubit) to restore the controlling ancilla qubit to the $\ket{0}$ state as shown in the last layer of the circuit in Figure~\ref{fig:oc2}.

\begin{figure}[htbp]
  \centering
  \begin{quantikz} 
     \lstick[1]{Validation qubit $\ket{v}$} & \targ{} & \qw\\ 
     \lstick[1]{Controlling qubit $\ket{0}$}& \qw & \qw \\
     \lstick[1]{Rotation qubit $\ket{0}$} &   \qw    & \qw \\
    \lstick[1]{Selection qubits $\ket{l}$}  &  \gate{C_l} & \qw \\
    \lstick[3]{Fock qubits $\ket{j}$} &  \ctrl{-4} & \qw \\
     &  \qw & \qw \\ 
     & \qw  & \qw 
\end{quantikz}
\caption{Illustration of $U_l$ circuit with $\mathcal{H}_l=\fc{a}{p}\fan{a}{p}$.}
\label{fig:oc3}
\end{figure}
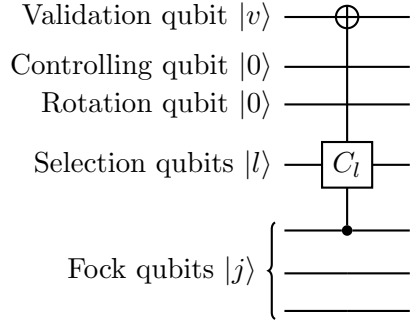

When $p=q$, no swap is needed when \eqref{eq:pqcond} holds. As a result, the corresponding $U_{l}$ circuit, which is shown in Figure~\ref{fig:oc3}, can be simplified. The construction is similar to encode the self energy term $\sum_{i} \fc{c}{i}\fan{c}{i}$ in the pairing Hamiltonian. 

To illustrate the cumulative effect of applying $(I\otimes I \otimes D_s \otimes I)(X\otimes I\otimes I\otimes I)(U_{L-1}\cdots U_1U_0)$ to the state $\ket{0}\ket{0}\ket{0}\ket{j}$, we consider a simple Hamiltonian of the form $\mathcal{H}=\mathcal{H}_0+\mathcal{H}_1$, which requires a single selection qubit in the $O_C$ circuit. We assume that both $(\ket{0},\ket{j})$ and $(\ket{1},\ket{j})$ satisfy \eqref{eq:pqcond}.  All intermediate quantum states produced from the application of each layer of the circuit shown in Figure~\ref{fig:OCsequence} (excluding the last $D_s$ layer) to the input state are shown below.
\begin{equation}
\label{equ:hsum}
  \begin{split}
    \ket{0} \ket{0} \ket{0} \ket{j} &
    \xrightarrow{D_s} \frac{1}{\sqrt{2}}\ket{0} \ket{0} \ket{0} \ket{j} + \frac{1}{\sqrt{2}}\ket{0}  \ket{0} \ket{1} \ket{j} \\
    &
    \xrightarrow{X} \frac{1}{\sqrt{2}}\ket{1} \ket{0}  \ket{0}  \ket{j} + \frac{1}{\sqrt{2}}\ket{1}  \ket{0}  \ket{1}  \ket{j} \\
    & \xrightarrow{U_0} \frac{1}{\sqrt{2}}\ket{0}  \ket{0}  \ket{0}  \ket{c(j,0)} + \frac{1}{\sqrt{2}}\ket{1}  \ket{0}  \ket{1}  \ket{j} \\
    & \xrightarrow{U_1} \frac{1}{\sqrt{2}}\ket{0}  \ket{0}  \ket{0}  \ket{c(j,0)} + \frac{1}{\sqrt{2}}\ket{0}  \ket{0}  \ket{1}  \ket{c(j,1)}
  \end{split}
\end{equation}
Taking the inner product of the last of line of \eqref{equ:hsum} and $(I\otimes I\otimes D_s \otimes I) \ket{0}\ket{0}\ket{0}\ket{i}$ results in
\begin{equation}
\bra{0}\bra{0}\bra{0}\bra{i} O_C \ket{0}\ket{0}\ket{0}\ket{j} 
= \frac{1}{2}\left[ \delta(i,c(j,0)) + \delta(i,c(j,1)) \right],
\label{eq:hij1}
\end{equation}
which correctly characterizes the nonzero structure of the $j$th column of $\mathcal{H}$.
We note that \eqref{equ:hsum} is independent of the sequence in which $U_l$'s acts.


If, for example, $(\ket{1},\ket{j})$ does not satisfy the condition \eqref{eq:pqcond}, then it follows that
\begin{equation}
\label{equ:hsum2}
  \begin{split}
    \ket{0}  \ket{0}  \ket{0}  \ket{j} & \xrightarrow{X,D_s} \frac{1}{\sqrt{2}}\ket{1}  \ket{0}  \ket{0}  \ket{j} + \frac{1}{\sqrt{2}}\ket{1}  \ket{0}  \ket{1}  \ket{j} \\
    & \xrightarrow{U_0} \frac{1}{\sqrt{2}}\ket{0}  \ket{0}  \ket{0}  \ket{c(j,0)} + \frac{1}{\sqrt{2}}\ket{1}  \ket{0}  \ket{1}  \ket{j} \\
    & \xrightarrow{U_1} \frac{1}{\sqrt{2}}\ket{0}  \ket{0}  \ket{0}  \ket{c(j,0)} + \frac{1}{\sqrt{2}}\ket{1}  \ket{0}  \ket{1}  \ket{j}.
  \end{split}
\end{equation}
In this case, the application of $U_1$ does not alter the second term on the right hand side of \eqref{equ:hsum2}. Taking the inner product of the last line of \eqref{equ:hsum2} with $(I\otimes I\otimes D_s \otimes I) \ket{0}\ket{0}\ket{0}\ket{i}$, we obtain
\begin{equation}
\bra{0}\bra{0}\bra{0}\bra{i} O_C \ket{0}\ket{0}\ket{0}\ket{j} 
= \frac{1}{2}\delta(i,c(j,0)).
\label{eq:hij2}
\end{equation}
Compared to \eqref{eq:hij1}, we see that the invalid operation drops out in \eqref{eq:hij2}. The calculation showed how $O_c$ can be constructed within a toy example. We summarize the result for more general application in the following theorem.
\begin{theorem}
\label{thm:Ublock}
  Suppose there exist oracles $U_i$ that works as 
  \begin{align}
    U_i : \ket{v} \ket{l} \ket{\ws}
    =\begin{cases}
      \ket{0} \ket{l} \ket{c(j,l)} & \quad \text{if $i=l, v=1$ and satisfies \eqref{eq:pqcond}} , \\
      \ket{v} \ket{l} \ket{j} & \quad \text{else} ,
    \end{cases}
  \end{align}
where $\ket{v} , \ket{l} ,\ket{j}$ respectively indicates the state on validation qubit, selection qubit and Fock qubit, then $O_C$ can be constructed as 
  \begin{equation}
    O_c = (X\otimes I)  U_0 U_1 \dots U_{L-1}
  \end{equation}
and the $O_c$ satisfies that
\begin{align}
    Oc : \ket{0} \ket{l} \ket{j} 
    \to \begin{cases}
      \ket{0} \ket{l} \ket{c(j,l)} & \quad \text{if satisfies \eqref{eq:pqcond}} , \\
      \ket{1} \ket{l} \ket{j} & \quad \text{else} .
    \end{cases}  
\end{align}


\end{theorem}
\begin{proof}
  The proof follows from direct calculations.
\end{proof}

\subsection{The $O_\mathcal{H}$ circuit}
\label{sec:Oh}
The $O_\mathcal{H}$ circuit is used to encode nonzero matrix elements through controlled rotations.  Since each nonzero element $v_l$ is associated with the coefficient of a particular $\mathcal{H}_{\ell}$ term in \eqref{eq:hpairnumber}, we construct a controlled rotation $ O_\mathcal{H}^{(l)} $ for each coefficient as 
\begin{equation}
  O_\mathcal{H}^{(l)} \ket{0} \ket{l} \ket{j}=(v_l\ket{0} + \sqrt{1-|v_l|^2}\ket{1} )\ket{l} \ket{j},
\end{equation}
when $l = (p,q)$) and $j$ satisfy \eqref{eq:pqcond}. 
  
\begin{figure}[htbp]
  \centering
  \begin{quantikz}
 \mbox{Rotation qubit}  & \ket{0} & \gate[3]{O_{\mathcal{H}}^{(l)}}  & \qw \\ 
 \mbox{Selection qubits}  & \ket{l} &  & \qw \\
 \mbox{Fock qubits}  & \ket{j} &  & \qw 
  \end{quantikz}
  =
  \begin{quantikz} 
     &\ket{0} & \gate[1]{R_{y}(\theta_{l})} & \qw \\ 
    \lstick[3]{$\ket{l}$}& \ket{l_0}  & \ctrl{-1} & \qw \\
    & \ket{l_1} & \octrl{-2}& \qw\\
    & \ket{l_2} & \ctrl{-1}& \qw\\
    &\ket{j} & \qw  & \qw
\end{quantikz} 
\caption{Illustration of $O_\mathcal{H}^{(l)}$ circuit for $l = \ket{101}$.}
\label{fig:oa2}
\end{figure}
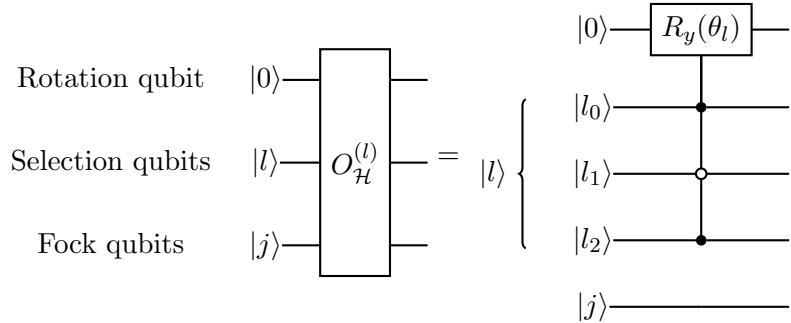

In the circuit implementation, we only need to place the control on the selection ancilla qubits $\ket{l}$.   It is not necessary to place additional controls on the system qubit $\ket{j}$ beacause if $l$ and $j$ do not satisfy \eqref{eq:pqcond}, the result of the controlled rotation will simply be discarded because the validation ancilla qubit will be in the $\ket{1}$ state after $O_C$ is applied to the output of $O_\mathcal{H}$.  Figure \ref{fig:oa2} gives an example of how $O_\mathcal{H}^{(l)}$ looks for some $l$ represented by $\ket{101}$.
The rotation $R_{y}(\theta_{l})$ is applied to the ``rotation qubit" with the rotation angle $\theta_l$ being $\arccos(v_l)$.

\begin{figure}[htbp]
  \centering
  \begin{quantikz}
    &\ket{0} & \gate[3]{O_\mathcal{H}}  & \qw \\ 
    &\ket{l}  & & \qw \\
    &\ket{j} & & \qw  
  \end{quantikz}=
  \begin{quantikz} 
     &\ket{0} & \gate[3]{O_{\mathcal{H}}^{(0)}} & \gate[3]{O_{\mathcal{H}}^{(1)}} & \qw \dots & \gate[3]{O_{\mathcal{H}}^{(L-1)}} & \qw \\ 
    &\ket{l}  & & & \qw \dots & \qw \dots & \qw \\
    &\ket{j} & \qw & \qw & \qw & \qw & \qw  
\end{quantikz} 
\caption{Illustration of $O_\mathcal{H}$ circuit.}
\label{fig:oa1}
\end{figure}
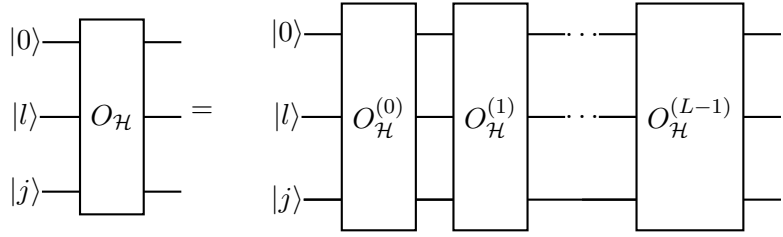

We note that the $(p,q)$ pairs associated with different terms in \eqref{eq:pairingModel_fermions} are mutually exclusive. Therefore, the $O_\mathcal{H}$ circuit can be implemented as a sequence of controlled rotations $O_\mathcal{H}^{(l)}$ ($l = 0, 1, 2, ..., L-1$) circuit components as shown in Figure \ref{fig:oa1}, i.e.,
\begin{equation}
  \begin{split}
    O_{\mathcal{H}} &=O_{\mathcal{H}}^{(L-1)}O_{\mathcal{H}}^{(L-2)}\cdots O_{\mathcal{H}}^{(1)} O_{\mathcal{H}}^{(0)} .  \end{split}
\end{equation}
We remark that $O_{\mathcal{H}}$ is independent of the order of instruction of $O_\mathcal{H}^{(l)}$.


\subsection{The complete circuit}

\begin{figure}
  \begin{center}
    \begin{quantikz}
    &\ket{0} \quad & \gate{X}           & \qw & \qw & \dots & \qw &\qw  &\gate[4]{U_{0}} & \dots & \gate[4]{U_{L-1}} &\qw &\qw \\ 
     &\ket{0} \quad & \qw          & \gate[2]{O_{\mathcal{H}}^{(0)}} & \gate[2]{O_{\mathcal{H}}^{(1)}} &\dots &\gate[2]{O_{\mathcal{H}}^{(L-1)}} &\qw   &\qw & \dots & \qw &\qw &\qw \\ 
    &\ket{l} \quad & \gate{D_s}     & \qw  & \qw       &\dots &\qw &\qw  &\qw &\dots &\qw & \gate{D_{s}} &\qw\\
    &\ket{j} \quad & \qw            & \qw & \qw        &\dots &\qw &\qw   &\qw &\dots &\qw & \qw &\qw
  \end{quantikz}
  \end{center}
  \caption{A schematic circuit design for the block encoding of $\mathcal{H}_{\rm pair}$.}
\label{fig:completeU}
\end{figure}
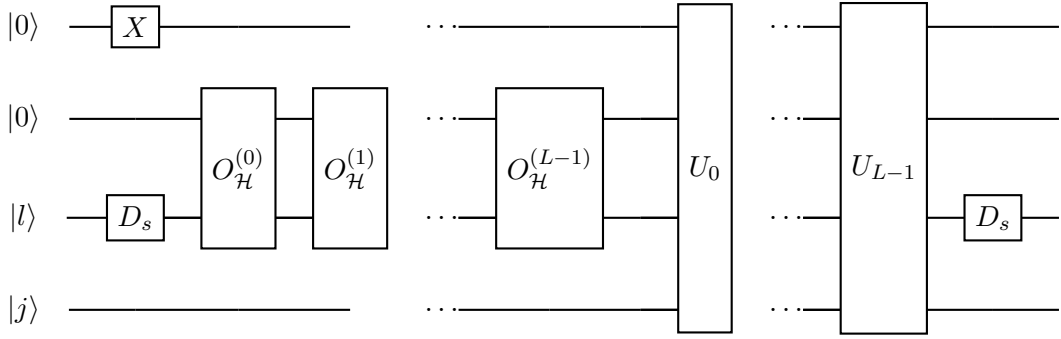

The complete circuit for block encoding $\mathcal{H}_{\rm pair}$ \eqref{eq:hpairnumber} is presented in Figure~\ref{fig:completeU}.

We now verify that such a circuit produces the desired output with the input to the circuit being
$\ket{0}\ket{0}\ket{0}\ket{0^m}\ket{j}$. 
In particular, we have 
\begin{equation}
\label{equ:circa}
\begin{split}
    \ket 0 \ket 0 \ket 0 \ket 0 \ket{0^m} \ket{j} &\xrightarrow{X\otimes D_s}\frac{1}{\sqrt{s}} \sum_{l=0}^{L-1} \ket 1 \ket 0 \ket{0} \ket{l} \ket{j} \\
    & \xrightarrow{O_\mathcal{H}} \frac{1}{\sqrt{s}} \sum_{l \in [s]} \ket{1} \ket{0} \left(v_l\ket{0} + \sqrt{1-|v_l|^2} \ket{1} \right) \ket{l} \ket{j} \\
    & \xrightarrow{O_C} \frac{1}{\sqrt{s}} \sum_{l \in [s]} \ket{o(j,l)} \ket{0} \left( v_l\ket{0} + \sqrt{1-|v_l|^2} \ket{1} \right) \ket{l} \ket{c(j,l)},
\end{split}
\end{equation}
with $[s] = 0, 1, \cdots , s-1$. We define $o(j,l)$ as
\begin{equation}
  o(j,l) = \begin{cases}
    0 &\text{if $l$ and $j$ satisfy \eqref{eq:pqcond}} , \\
    1 &\text{otherwise}.
  \end{cases}
\end{equation}
By taking the inner product of the output of \eqref{equ:circa} with $(I\otimes I \times I \otimes D_s \otimes I)\ket{0}\ket{0}\ket{0}\ket{o^m}\ket{i}$, 
we obtain
\begin{equation}
    \bra{0} \bra{0} \bra{0} \bra{0^m} \bra{i} U_{\mathcal{H}} \ket 0 \ket 0 \ket 0 \ket{0^m}\ket{j}=\frac{1}{s} v_l \delta(i,c(j,l)),
\end{equation}
when $o(j,l)=0$. Otherwise, this matrix element is 0 due to the vanishing inner product between $\ket{0}$ and $\ket{1}$ of the validation qubit.

\subsection{Gate complexity}
\label{sec:GateComplexity}
{
The complete block encoding circuit shown in Figure \ref{fig:completeU} contains $L$ $U_{\ell}$-blocks, where $L$ is the number of creation and annihilation pairs in \eqref{eq:Ham1}. Each one of these blocks consists of at most one $\log(L)+3$ qubit CNOT gate and one single qubit controlled swap gate as well as two standard CNOT gates as we have shown in section~\ref{sec:circuit_for_U}. The multi-qubit control gate can be decomposed into $2\log(L)+3$ three-qubit Toffoli gates~\cite{vale2023decomposition}. Each three-qubit Toffoli gate can in turn be decomposed into seven T gates and six CNOT gates~\cite{shende2008cnot}. The control swap in each $U_\ell$ block can be implemented as two standard CNOTs. Therefore, in total, each $U_{\ell}$ block requires $12\log(L)+22$ CNOT gates and $14\log(L)+21$ T gates. Assume each controlled rotation gate can be directly implemented and each such gate is counted as one two-qubit gate, then in the complete block encoding circuit, there are $L$ two-qubit gates for controlled rotation. Adding up all these gate counts, we estimate the circuit requires $12L\log(L)+23L$ two-qubit gates and $14L\log(L)+21L$ T gates. Because $L$ is generally a polynomial of $n = \log N$, the overall  gate complexity of the circuit is $\mathcal{O}(\text{poly}(n)\log(n))$.
We should note that our estimation here assumes the quantum computer we work with has all-to-all connectivity and can perform an arbitrary single-qubit rotation for any angle.  In practice, these assumption may not hold for the current generation of noisy intermediate-scale quantum (NISQ) devices.  A more detailed analysis of the gate count for a particular quantum computer is beyond the scope of this paper.
}

\section{Numerical examples}
\label{sec:examples}
In this section, we give two examples of explicit circuit design for second-quantized Hamiltonians. The first examples involves a toy Hamiltonian with $4$ (pseudo) one-body terms.  The second example consists of a Hamiltonian constructed in a Fock space defined by a limited number of single-particle states.
\subsection{$O_C$ for a simple example}
\begin{figure}[htbp]
  \centering
\begin{quantikz} 
     \lstick[1]{$\ket{v}$} & \targ{} & \qw\\ 
     \lstick[1]{$\ket{0}$}& \qw & \qw \\
     \lstick[1]{$\ket{0}$} &   \qw    & \qw \\
    \lstick[1]{$\ket{l_0}$}  &  \ctrl{-3} & \qw \\
    \lstick[1]{$\ket{l_1}$}  &  \ctrl{-2} & \qw \\
    \lstick[1]{$\ket{j_0}$} &  \qw & \qw \\
    \lstick[1]{$\ket{j_1}$}  &  \ctrl{-2} & \qw 
\end{quantikz}
     \quad
     \begin{quantikz} 
     \lstick[1]{$\ket{v}$} & \ctrl{3} & \qw & \targ{}  & \octrl{1} & \qw\\ 
     \lstick[1]{$\ket{0}$} &   \targ{}  & \ctrl{4} & \ctrl{-1}   & \targ{} & \qw \\
          \lstick[1]{$\ket{0}$} &\qw &\qw &\qw  &\qw & \qw \\
    \lstick[1]{$\ket{l_0}$}  &  \octrl{1} & \qw & \qw & \qw  & \qw \\
    \lstick[1]{$\ket{l_1}$}  &  \ctrl{1} & \qw & \qw & \qw  & \qw \\
    \lstick[1]{$\ket{j_0}$} &  \ctrl{1} & \swap{1} & \qw & \qw & \qw \\
    \lstick[1]{$\ket{j_1}$} &  \octrl{} & \swap{}   & \qw & \qw &\qw
     \end{quantikz}

\caption{The left panel shows $U_3$ circuit with $\sH_3=\fc{a}{1}\fan{a}{1}$ and the right panel shows $U_2$ circuit with $\sH_2=\fc{a}{1}\fan{a}{0}$.}
\label{fig:u23new}
\end{figure}
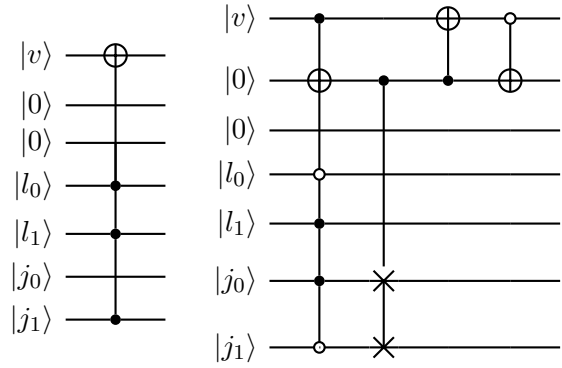

In this section, we will examine how a block encoding circuit can be constructed for the following simple model Hamiltonian 
\begin{equation} \sH=\underbrace{\fc{a}{0}\fan{a}{0}}_{\sH_0}+\underbrace{\fc{a}{0}\fan{a}{1}}_{\sH_1}+\underbrace{\fc{a}{1}\fan{a}{0}}_{\sH_2}+\underbrace{\fc{a}{1}\fan{a}{1}}_{\sH_3}, \label{eq:hsimple}
\end{equation}
using the techniques discussed above. Since the coefficient for each $\sH_l$ in \eqref{eq:hsimple} is $1$, we only need to examine the $O_C$ circuit as $O_\sH$ is simply an identity. Each creation and annihilation operator is for a pair of nucleons and due to the form of the Hamiltonian, only the states with occupied nucleon pairs will be nonzero. It suffice to only encode the Hamiltonian for the states where nucleons appear in pairs as all other states will vanish under the Hamiltonian. Because there are four terms in the Hamiltonian, we use $l=0,1,2,3$ to index each term, and map the binary representation of $l=l_0 + 2 l_1$ with $l_0,l_1 \in \{0,1\}$ to the ancilla qubits labelled by $\ket{l_0}$ and $\ket{l_1}$ in Figure~\ref{fig:u01new}.
Furthermore, since there are only two particles involved in the Hamiltonian, we use two system qubits labeled as $\ket{j_0}$ and $\ket{j_1}$ with $j_0,j_1 \in \{0,1\}$ to represent a computational basis in the 4-dimensional Fock space associated with the Hamiltonian $\sH$ defined in \eqref{eq:hsimple}. Figure \ref{fig:u01new} and Figure \ref{fig:u23new} illustrate the explicit construction of each $U$ block defined in \cref{thm:Ublock} for the Hamiltonian \eqref{eq:hsimple}.

\begin{figure}[htbp]
  \centering
\begin{quantikz} 
     \lstick[1]{$\ket{v}$} & \targ{} & \qw\\ 
     \lstick[1]{$\ket{0}$}& \qw & \qw \\
     \lstick[1]{$\ket{0}$} &   \qw    & \qw \\
    \lstick[1]{$\ket{l_0}$}  &  \octrl{-3} & \qw \\
    \lstick[1]{$\ket{l_1}$}  &  \octrl{-1} & \qw \\
    \lstick[1]{$\ket{j_0}$} &  \ctrl{-1} & \qw \\
     \lstick[1]{$\ket{j_1}$} &  \qw & \qw 
\end{quantikz}
     \quad
     \begin{quantikz} 
     \lstick[1]{$\ket{v}$} & \ctrl{4} & \qw & \targ{}  & \octrl{1} & \qw\\ 
     \lstick[1]{$\ket{0}$} &   \targ{}  & \ctrl{4} & \ctrl{-1}   & \targ{} & \qw \\
          \lstick[1]{$\ket{0}$} &\qw &\qw &\qw  &\qw & \qw \\
    \lstick[1]{$\ket{l_0}$}  &  \ctrl{} & \qw & \qw & \qw  & \qw \\
    \lstick[1]{$\ket{l_1}$}  &  \octrl{1} & \qw & \qw & \qw  & \qw \\
    \lstick[1]{$\ket{j_0}$} &  \octrl{1} & \swap{1} & \qw & \qw & \qw \\
    \lstick[1]{$\ket{j_1}$} &  \ctrl{} & \swap{}   & \qw & \qw &\qw
     \end{quantikz}

\caption{The left panel shows $U_0$ circuit with $\sH_0=\fc{a}{0}\fan{a}{0}$ and the right panel shows $U_1$ circuit with $\sH_1=\fc{a}{0}\fan{a}{1}$.}
\label{fig:u01new}
\end{figure}
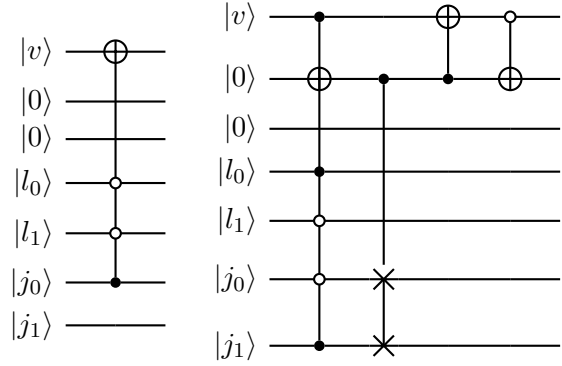

\begin{figure}[htbp]
  \centering
\begin{quantikz} 
     \lstick[1]{$\ket{1}$} & \ctrl{4} \slice{1} & \qw^{\ket{1}}& \qw \slice{2} &\qw & \targ{} \slice{3} &\qw^{\ket{0}}  & \octrl{1} \slice{4} & \qw^{\ket{0}} \\
     \lstick[1]{$\ket{0}$} &   \targ{} &\qw^{\ket{1}} & \ctrl{4} &\qw & \ctrl{-1}  &\qw & \targ{} & \qw^{\ket{0}} \\
          \lstick[1]{$\ket{0}$} &\qw &\qw &\qw &\qw  &\qw &\qw &\qw & \qw \\
    \lstick[1]{$\ket{1}$}  &  \ctrl{} & \qw & \qw &\qw & \qw &\qw & \qw  & \qw \\
    \lstick[1]{$\ket{0}$}  &  \octrl{1} &\qw  & \qw &\qw & \qw &\qw & \qw  & \qw \\
    \lstick[1]{$\ket{0}$} &  \octrl{1} &  \qw & \swap{1} &\qw^{\ket{1}} & \qw  &\qw & \qw & \qw^{\ket{1}} \\
     \lstick[1]{$\ket{1}$} &  \ctrl{} &\qw & \swap{}   & \qw^{\ket{0}} & \qw & \qw & \qw &\qw^{\ket{0}}
     \end{quantikz}

\caption{Illustration of the action of $U_1$ on the state $\ket{1}\ket{l}\ket{\ws}$ with $l=1$ and $ \ket{ j } = \fc{a}{1}\ket{vac}$}
\label{fig:udetailnew}
\end{figure}
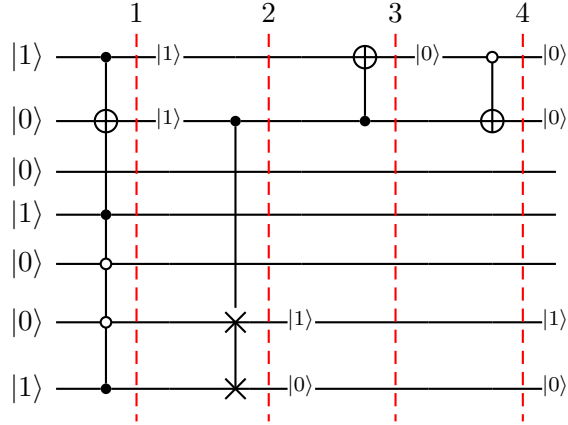

To illustrate the effect of each layer of the circuit, we show the input to the quantum circuit, the intermediate state produced by each layer of the $O_C$ circuit, and the output state in Figure~\ref{fig:udetailnew}. In step 1, a multi-qubit control is used to ensure the controlling qubit is turned into $\ket{1}$ only when the three conditions listed in Theorem \ref{thm:Ublock} are satisfied. If any one of them is violated, then the first layer of the circuit does nothing to the input state. In step 2, a control swap operation is applied under condition of the controlling qubit being $\ket{1}$. This indicates the action of $\mathcal{H}_{1} = a_0^{\dagger}a_1$ on the Fock state when all the three conditions are satisfied. In step 3, a CNOT is applied to the validation qubit, in order to record that the output is valid and should be saved for final computation. In step 4, another CNOT is applied to uncompute the controlling qubit.

\subsection{The three-nucleon system with the pairing Hamiltonian}
\label{sec:threenucleon}

In this subsection, we present the circuit to block encode a simple pairing Hamiltonian $\sH_{\rm pair}$ \eqref{eq:pairingModel_fermions} and possible applications based on our block encoding scheme. Instead of block encoding the Hamiltonian with one-body pseudo-particle creation and annihilation  operators, we illustrate how to extend the circuit construction according to \eqref{eq:pairingModel_fermions} with same idea discussed above. In particular, we elect to demonstrate our scheme with the three-nucleon problem in Ref. \cite{Du:2023bpw}. For this model problem, we choose a set of six single-particle bases 
shown in Table \ref{tab:basisPairingHamiltonian}. Each single particle basis is labeled by a set of quantum numbers $ { \{ } n_{\alpha}$, $l_{\alpha}$, $j_{\alpha}$, $m_{j_{\alpha}} { \} } $, where $m_{j_{\alpha}} = \pm m_{\alpha}$ as we explained in section~\ref{sec:intro}.  
We omit the spin, which is taken to be $1/2$ for the nucleons. In our demonstration, we do not distinguish between protons and neutrons; hence we also omit the isospin degree of freedom in our basis. With the mapping defined in Table~\ref{tab:basisPairingHamiltonian}, we can label each single particle basis by an integer listed in the first column of  Table~\ref{tab:basisPairingHamiltonian} and  rewrite the pairing Hamiltonian as
\begin{align}
\label{equ:H}
  \begin{split} 
\sH_{pair} = & c_0^{\dagger}c_1^{\dagger}c_{1}c_{0} +  c_0^{\dagger}c_1^{\dagger}c_{3}c_{2}+c_0^{\dagger}c_1^{\dagger}c_{5}c_{4}\\ 
&+c_2^{\dagger}c_3^{\dagger}c_{1}c_{0}+c_2^{\dagger}c_3^{\dagger}c_{3}c_{2}+c_2^{\dagger}c_3^{\dagger}c_{5}c_{4}\\ 
&
+c_4^{\dagger}c_5^{\dagger}c_{1}c_{0}+c_4^{\dagger}c_5^{\dagger}c_{3}c_{2}+c_4^{\dagger}c_5^{\dagger}c_{5}c_{4} . \\
\end{split} 
\end{align}

\begin{table}[!ht] 
\caption{The restricted single-particle basis set for a single-species three-nucleon system. See text for details.}
\centering
\begin{tabular}{c | c c c c }
\hline \hline
single-particle basis   &  & quantum numbers &  &   
\\
(qubit) index  & $\ n\ $ & $\ l\ $ & $\ 2j\ $ & $\ 2m_j\ $  \\ 
\hline
${\bf 0} $ &  $0 $ & $0 $ & $1$ & $-1$  \\ 
${\bf 1} $ &  $0 $ & $0 $ & $1$ & $+1$  \\ 
${\bf 2} $ &  $1 $ & $0 $ & $1$ & $-1$  \\ 
${\bf 3} $ &  $1 $ & $0 $ & $1$ & $+1$  \\ 
${\bf 4} $ &  $2 $ & $0 $ & $1$ & $-1$  \\ 
${\bf 5} $ &  $2 $ & $0 $ & $1$ & $+1$  \\ 
\hline \hline
\end{tabular} 
\label{tab:basisPairingHamiltonian}
\end{table}

\subsubsection{Construction of $O_C$ oracle}
As the coefficient for each term in the Hamiltonian given by \eqref{equ:H} is $1$, it is sufficient to consider only the $O_C$ circuit for the block encoding of $\sH_{pair}$. Instead of encoding a neucleon pair with a qubit, we construct the block encoding with the direct encoding scheme of the single-particle states.
In the Hamiltonian, there are in total 9 pairing terms where each term can be labeled by a pair of indices $(l_1,l_2)$ that specify the pairing term, 

\begin{equation}
c_{2l_1}^{\dagger}c_{2l_1+1}^{\dagger}c_{2l_2+1}c_{2l_2}
\end{equation}
 where $l_1, l_2 \in \{ 0, 1,2 \}$. 
 
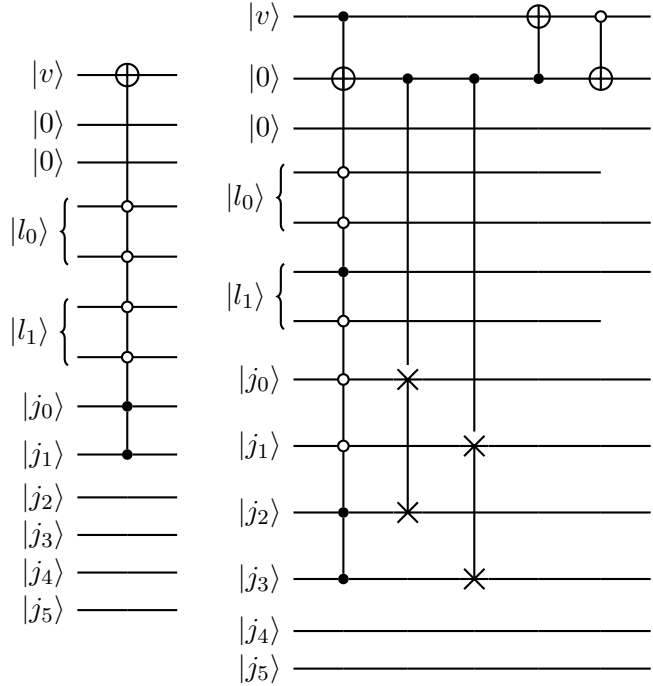
\begin{figure}[htbp]
  \centering
\begin{quantikz} 
     \lstick[1]{$\ket{v}$} & \targ{} & \qw\\ 
     \lstick[1]{$\ket{0}$}& \qw & \qw \\
     \lstick[1]{$\ket{0}$} &   \qw    & \qw \\
    \lstick[2]{$\ket{l_0}$}  &  \octrl{-3} & \qw \\
      &  \octrl{-1} & \qw \\
        \lstick[2]{$\ket{l_1}$}  &  \octrl{-1} & \qw \\
      &  \octrl{-1} & \qw \\
    \lstick[1]{$\ket{j_0}$} &  \ctrl{-1} & \qw \\
     \lstick[1]{$\ket{j_1}$} &  \ctrl{-1} & \qw \\
          \lstick[1]{$\ket{j_2}$} &  \qw & \qw \\
               \lstick[1]{$\ket{j_3}$} &  \qw & \qw \\
                    \lstick[1]{$\ket{j_4}$} &  \qw & \qw \\
                       \lstick[1]{$\ket{j_5}$} &  \qw & \qw 
\end{quantikz}
     \quad
     \begin{quantikz} 
     \lstick[1]{$\ket{v}$} & \ctrl{3} & \qw & \qw & \targ{}  & \octrl{1} & \qw\\ 
     \lstick[1]{$\ket{0}$} &   \targ{}  & \ctrl{6} &\ctrl{7} & \ctrl{-1}   & \targ{} & \qw \\
          \lstick[1]{$\ket{0}$} &\qw &\qw &\qw &\qw  &\qw & \qw \\
    \lstick[2]{$\ket{l_0}$}  &  \octrl{1} & \qw & \qw & \qw  & \qw \\
      &  \octrl{1} & \qw &\qw & \qw & \qw  & \qw \\
        \lstick[2]{$\ket{l_1}$}  &  \ctrl{1} & \qw  &\qw & \qw & \qw  & \qw \\
     &  \octrl{1} & \qw & \qw & \qw  & \qw \\
    \lstick[1]{$\ket{j_0}$} &  \octrl{1} & \swap{2} &\qw & \qw & \qw & \qw \\
    \lstick[1]{$\ket{j_1}$} &  \octrl{1} & \qw  &\swap{2}  & \qw & \qw &\qw \\
     \lstick[1]{$\ket{j_2}$} &  \ctrl{1} & \swap{} &\qw &\qw &\qw &\qw \\
               \lstick[1]{$\ket{j_3}$} &  \ctrl{} &\qw & \swap{} &\qw &\qw  &\qw \\
                    \lstick[1]{$\ket{j_4}$} &  \qw &\qw & \qw &\qw &\qw &\qw \\
                       \lstick[1]{$\ket{j_5}$} &  \qw &\qw & \qw &\qw &\qw &\qw
     \end{quantikz}

\caption{The left panel shows circuit of $U_{0,0}$ corresponds to $c_0^{\dagger}c_1^{\dagger}c_{1}c_{0}$ and the right panel shows $U_{0,1}$ circuit with $c_0^{\dagger}c_1^{\dagger}c_{3}c_{2}$.}
\label{fig:u3}
\end{figure}

The $O_C$ circuit can be constructed via a sequence of $U_{l_1,l_2}$ blocks. We illustrate the explicit construction in the \cref{fig:u3}. We still follow our circuit design in previous section but explicitly encode the state of each neucleon clearly, i.e, $\ket{j_0}, \ket{j_1},\ket{j_2},\ket{j_3},\ket{j_4},\ket{j_5}$ each represents neucleon is occupied or not. The condition on $l_1$ and $l_2$ is transformed into the multi-control of binary representation of $l_1$ and $l_2$. The condition of \ref{eq:pqcond} is equivalent to the the control of 
$\ket{j_0,j_1}=\ket{00}$ and $\ket{j_2,j_3}=\ket{11}$. The swap operation in the formalism is the swap between the state $\ket{j_0,j_1} $ and $\ket{j_2, j_3}$ and it could be implemented as four CNOT gate in practice.

\subsubsection{Verification of block encoding}
In this subsection, we validate the block encoding circuit by evaluating the output of the circuit when a particular computational basis is chosen as the input. For this three-nucleon system in the restricted single-particle basis set (Table \ref{tab:basisPairingHamiltonian}), we have $\tbinom{6}{3}=20$ three-nucleon states in total yielding a $20\times 20$ Hamiltonian matrix. The Hamiltonian can be block diagonalized and partitioned according to the $M_J$ value of a many-body state. In particular, there are 1) 9 states with $M_J = -\frac{1}{2}$; 2) 9 states with $M_J = +\frac{1}{2}$; 3) 1 state with $M_J = -\frac{3}{2}$; and 4) 1 state with $M_J = +\frac{3}{2}$. We sort these three-nucleon states in Table \ref{tab:mJ_scheme_pairing_model}. The states associated with a distinct $M_J$ value form an invariant subspace of $\sH_{\text{pair}}$. For this problem, we can evaluate the matrix elements of the pairing Hamiltonian within each diagonal block associated with a distinct $M_J$ value analytically. The many-nucleon basis states that span each invariant subspace associated with a distinct $M_J$ value are shown in Table \ref{tab:mJ_scheme_pairing_model}. The diagonal block associated with $M_J=+\frac{1}{2}$, which corresponds to the invariant subspace spanned by the states
\begin{align*}
     ( 0,1,3 ) , ( 0,1,5 ) , ( 0,3,5 ) , ( 1,2,3 ) , ( 1,2,5 ) , ( 1,3,4 ) , ( 1,4,5 ) , ( 2,3,5 ) , ( 3,4,5 ) , 
\end{align*}
for example, has the form
\begin{align}
\sH_{\rm pair}\big(M_J= + \frac{1}{2} \big) = 
\begin{pmatrix}
1 & 0 & 0 & 0 & 0 & 0 & 0 & 0 & 1 \\ 
0 & 1 & 0 & 0 & 0 & 0 & 0 & 1 & 0 \\ 
0 & 0 & 0 & 0 & 0 & 0 & 0 & 0 & 0 \\ 
0 & 0 & 0 & 1 & 0 & 0 & 1 & 0 & 0 \\ 
0 & 0 & 0 & 0 & 0 & 0 & 0 & 0 & 0 \\ 
0 & 0 & 0 & 0 & 0 & 0 & 0 & 0 & 0 \\ 
0 & 0 & 0 & 1 & 0 & 0 & 1 & 0 & 0 \\ 
0 & 1 & 0 & 0 & 0 & 0 & 0 & 1 & 0 \\ 
1 & 0 & 0 & 0 & 0 & 0 & 0 & 0 & 1
\end{pmatrix}. \label{eq:matrix_rep_pairing_Hamiltonian}
\end{align}

\begin{table*}[!ht] 
\caption{The three-nucleon states sorted according to $M_J$. Only the indices of the occupied single-particle states are recorded, where the quantum numbers of each single-particle state are shown in Table \ref{tab:basisPairingHamiltonian}. For example, the three-nucleon state $(0,1,3)$ is equivalent to $|110100 \rangle $ in notations.}
\begin{tabular}{cccccccccc}
\hline \hline
$M_J$  & \multicolumn{9}{c}{three-nucleon state}               \\
\hline
$+3/2$ & $\ (1,3,5) \ $ &  --  &  --  &  --  &  --   &   -- &   -- &  --   &  --   \\
$+1/2$ & $\ (0,1,3) \ $ & $\ (0,1,5) \ $ & $\ (0,3,5) \ $ & $\ (1,2,3) \ $ & $\ (1,2,5) \ $ & $\ (1,3,4) \ $ & $\ (1,4,5) \ $ & $\ (2,3,5) \ $ & $\ (3,4,5) \ $ \\
$-1/2$ & $\ (0,1,2) \ $ & $\ (0,1,4) \ $ & $\ (0,2,3) \ $ & $\ (0,2,5) \ $ & $\ (0,3,4) \ $ & $\ (0,4,5) \ $ & $\ (1,2,4) \ $ & $\ (2,3,4) \ $ & $\ (2,4,5) \ $ \\
$-3/2$ & $\ (0,2,4) \ $ &  --  &  --  &  --  &  --   &   --  &  --   & --   & --  \\ \hline  \hline
\end{tabular}
\label{tab:mJ_scheme_pairing_model}
\end{table*}

We now verify the correctness of the block encoding circuit by examining the output of the circuit when the input of the circuit is initialized as $\ket{110100}$ which corresponds to the state $(0,1,3)$. This state is mapped to the first column $\sH_{\text{pair}}(M_J=+\frac{1}{2})$ defined in \eqref{eq:matrix_rep_pairing_Hamiltonian}.

\begin{equation}
\label{equ:blockv}
\begin{split}
    \ket 0  \ket{0^2} \ket{0^2} \ket{110100} &\xrightarrow{X\otimes D_s}\frac{1}{\sqrt{16}} \sum_{l_1=0}^{3}\sum_{l_2=0}^{3} \ket 1 \ket{l_1} \ket{l_2} \ket{110100} \\
    & \xrightarrow{O_\sH=I} \frac{1}{\sqrt{16}} \sum_{l_1=0}^{3}\sum_{l_2=0}^{3} \ket 1 \ket{l_1} \ket{l_2} \ket{110100} \\
    & \xrightarrow{U_{0,0}} \frac{1}{\sqrt{16}}  \ket{0} \ket{0} \ket{0} \ket{110100} + \frac{1}{\sqrt{16}}\sum_{(l_1, l_2) \neq (0,0)} \ket 1 \ket{l_1} \ket{l_2} \ket{110100}, \\
     & \xrightarrow{U_{0,1}} \frac{1}{\sqrt{16}}  \ket{0} \ket{0} \ket{0} \ket{110100} +  \frac{1}{\sqrt{16}}\ket 1 \ket{0} \ket{1} \ket{110100}, \\
     & +\frac{1}{\sqrt{16}}\sum_{(l_1, l_2) \neq (0,0), (0,1)} \ket 1 \ket{l_1} \ket{l_2} \ket{110100} \\
     & \dots \dots \\ 
     & \xrightarrow{U_{2,2}} \frac{1}{\sqrt{16}}  \ket{0} \ket{0} \ket{0} \ket{110100} + \frac{1}{\sqrt{16}}  \ket{0} \ket{0} \ket{2} \ket{000111}+ \frac{\sqrt{14}}{\sqrt{16}} \ket{1} \ket{\Psi} \ket{110100}\\
\end{split}
\end{equation}
where $\Psi = \sum_{(l_1,l_2)\neq(0,0),(0,1)} \ket{l_1}\ket{l_2}$. 
Taking the inner product of the last line of \eqref{equ:blockv} with $(I\otimes D_s \otimes I) \ket{\phi}$ results in 0 unless
$\ket{\phi} = \ket{110100}$ or $\ket{000111}$.  These two states correspond to the first and last rows of the matrix given in \eqref{eq:matrix_rep_pairing_Hamiltonian}. For these two states, the inner product yields 1/16. Similar verification can be made for other input states that correspond exactly to other columns of the matrix \eqref{eq:matrix_rep_pairing_Hamiltonian}, which ultimately shows that the circuit constructed above indeed produces a (16,5)-block encoding of $\sH_{\rm pair}\big(M_J= + \frac{1}{2} \big)$.

\subsection{The density of states via the block-encoded pairing Hamiltonian}
\label{sec:computedos}
In the section, we utilize the block encoding circuit of the pairing Hamiltonian to construct an explicit circuit for approximating the density of states (DOS) of the scaled pairing Hamiltonian $\mathcal{H}'_{\rm pair}$ for all the many-body systems allowed by the restricted basis set. This example serves to demonstrate an application of the block encoding. The DOS of the pairing Hamiltonian $\sH _{\rm pair}$ within the Hilbert space defined by all allowed fermion occupancies of the $n=6$ single-particle states described in section~\ref{sec:threenucleon} can be formally written as 
\begin{equation}
    d(\omega) \equiv \frac{1}{N}\mathrm{Tr}(\delta(\mathcal{H}'_{\rm pair} -\omega))=\frac{1}{N}\sum_{j=0}^{N-1}\delta(\lambda_{j}-\omega),
\end{equation}
where $N = 2^n = 2^6=64$ is the dimension of this Hilbert space,  
and $\lambda_{i}$ is $i$-th eigenvalue of $\sH '_{\rm pair}$. 

To approximate $d(\omega)$ computationally, one often replaces the Dirac-$\delta$ distribution by a smooth surrogate function such as a Gaussian or a Lorentzian~\cite{lin2016approximating}. If we use a Gaussian defined as $g(\omega) = \exp(-\frac{\omega^2} {2\sigma^2})$ to replace the Dirac-$\delta$ distribution, $d(\omega)$ becomes
\begin{equation} 
\begin{split} 
    d(\omega) \approx \frac{1}{N}\Tr(g( \mathcal{H}'_{\rm pair} -\omega I)) = \frac{1}{N} \sum_{j=0}^{N-1} g(\lambda_j -\omega).
\end{split} 
\label{eq:tracedos}
\end{equation} 

One way to develop a quantum algorithm to estimate $d(\omega)$ is to construct quantum circuits that can produce  $\ket{0^a}\ket{\psi_i}+\ket{0^a}^{\perp}\ket{\phi_i}$, where $\ket{\psi_i} = \frac{1}{\sqrt{N}}\sum_{x=0}^{N-1} \sqrt{g(\sH-\omega_i I)} \ket{x}\ket{x}$ 
for a set of sampled energy levels $\omega_i$, $i=1,2,...,m$ and $a$ is the number of ancilla qubits, $\ket{0^a}^{\perp}$ denotes a quantum state orthogonal to $\ket{0^a}$ and $\ket{\phi_i}$ is a quantum state that we do not care about. Here $\ket{x}$ denotes an $n$-qubit computational basis where $n = \log N$.  The success probability of measuring the ancilla qubits in the $\ket{0}^a$ state is $\braket{\psi_i}{\psi_i}$, which is exactly $d(\omega_i)$ as
\begin{align}
\braket{\psi_i}{\psi_i} &=  
\frac{1}{N} \sum_{x,y} [\sqrt{g( \mathcal{H}'_{\rm pair} -\omega_i I)}  \ket{y} \ket{y}]^{\dagger} \sqrt{g(\mathcal{H}'_{\rm pair} -\omega_i I)}  \ket{x} \ket{x}  \nonumber \\
&=\frac{1}{N} \sum_{x,y} \langle y | g( \mathcal{H}'_{\rm pair} -\omega_i I) | x \rangle \braket{y}{x} \nonumber \\
&=\frac{1}{N} \sum_{x} \langle x | g( \mathcal{H}'_{\rm pair} -\omega_i I) | x \rangle \nonumber \\
&=\frac{1}{N} \text{Tr} \left[ g( \mathcal{H}'_{\rm pair} -\omega_i I) \right].
\label{eq:psinorm2}
\end{align}

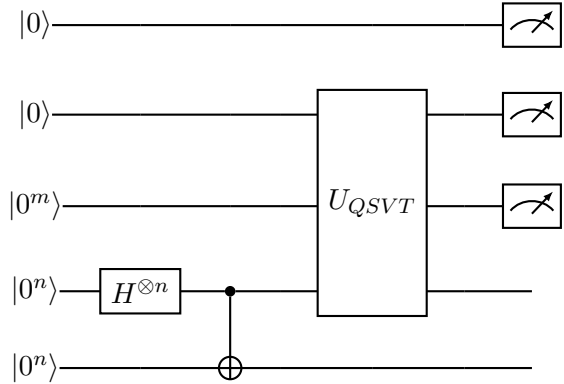
\begin{figure}[htbp]
\centering
  \begin{quantikz} 
    \ket{0}   &  \qw& \qw & \qw & \qw                 & \qw & \meter{} \\ 
    \ket{0} & \qw &  \qw & \qw & \gate[3]{U_{QSVT}}{} & \qw & \meter{} \\
    \ket{0^{m}} &\qw & \qw & \qw &                     & \qw & \meter{} \\
    \ket{0^{n}} & \gate[1]{H^{\otimes n}}{} &  \ctrl{1} & \qw &                     & \qw & \qw \\
    \ket{0^{n}} &  \qw & \targ{}  & \qw & \qw                 & \qw & \qw \\
\end{quantikz}
\caption{The overall structure of the circuit used to produce $\ket{0}^a\ket{\psi_i}$. The circuit block labelled by $U_{QSVT}$ is the circuit that block encodes $\sqrt{g( \mathcal{H}'_{\rm pair} -\omega_i I)}$.  The superposition of all $n$-qubit computational basis $\ket{\chi} = \frac{1}{\sqrt{N}}\sum_{x} \ket{x}\ket{x}$ is prepared via the Hadamard gates and multi-qubit CNOT gate (Figure 16).}
\label{fig:doscircuit}
\end{figure}

To produce $\ket{0^a}\ket{\psi_i}+\ket{0^a}^{\perp}\ket{\phi_i}$, we construct a circuit that block encodes $\sqrt{g( \mathcal{H}'_{\rm pair} -\omega_i I)}$.  This construction can be achieved by using QSP (Theorem~\ref{thm:qsp}) and QSVT.  We label this circuit as $U_{QSVT}$ in Figure~\ref{fig:doscircuit} which presents the overall circuit structure used to produce $\ket{0^a}\ket{\psi_i}$ for a fixed $\omega_i$.

\begin{figure}[htbp]
\centering
  \begin{quantikz} 
    \ket{0} &\gate{H}& \targ{}& \gate[1]{e^{-i\phi_{d}Z}}{} & \targ{}& \qw & \targ{}& \gate[1]{e^{-i\phi_{d-1}Z}}{}&\targ{} &\qw &\qw \dots & \gate{H} &\qw \\ 
     \ket{0^{m}} & \qw &\octrl{-1} & \qw & \octrl{-1}&\gate[2]{U_{B}} & \octrl{-1} & \qw & \octrl{-1} & \gate[2]{U_{B}^{\dagger}} & \qw \dots  & \qw &\qw \\
    \ket{\psi}  &  \qw & \qw &\qw & \qw & \qw &\qw &\qw & \qw &\qw &\qw \dots &\qw &\qw \\
\end{quantikz}
\caption{The circuit structure of the quantum singular value transformation. The phases are computed through QSP-PACK \cite{Dongefficient,Wang2022energylandscapeof,dong2022infinite} for the function $\sqrt{\tilde{g}}(\lambda-\omega_j)= \exp{\frac{(|\lambda|-\omega_j)^2}{4\sigma^2}}$ with $\sigma$ is set to be 0.01. When the QSP-PACK is used in the numerical experiments, we set the scaling factor to be 0.8, the maximum iteration of the optimization to be 200, and the precision to be $10^{-12}$. 
We use the Newton method for the optimization and set two options  \textit{targePre} and \textit{useReal} in the QSP-PACK to be true and false, respectively.  
}
\label{fig:QSP_example}
\end{figure}
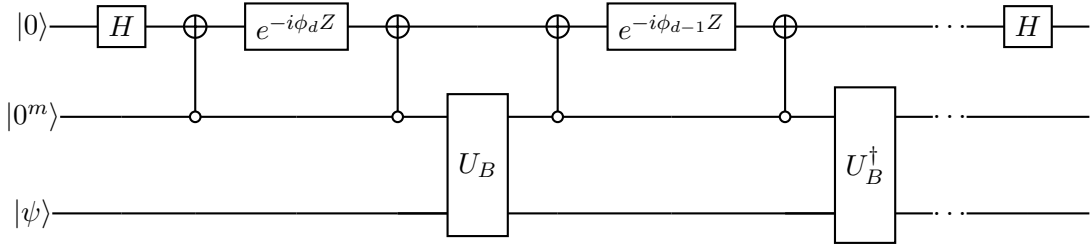

The general structure of the $U_{QSVT}$ circuit is shown in Figure~\ref{fig:QSP_example} where the circuit block labelled by $U_B$ represents the block encoding circuit for $ \mathcal{H}'_{\rm pair} $, and the phase angle $\phi_j$ that appears in a single qubit gate $e^{-i\phi_jZ}$ can be obtained from the software package QSP-PACK~\cite{Dongefficient,Wang2022energylandscapeof,dong2022infinite}. These angles are computed by solving an optimization problem that minimizes the difference between $\sqrt{g(\lambda-\omega_j)}$ and a polynomial approximation $p(\lambda)$. The optimization problem can be simplified for even and odd functions. Therefore, to take advantage of this simplification, we modify the surrogate Gaussian function by replacing $\lambda$ with $|\lambda|$ and obtain
\[
\tilde{g}(\lambda-\omega_j)= \exp{\frac{(|\lambda|-\omega_j)^2}{2\sigma^2}}.
\]
 
 \begin{figure}[htbp]
\centering
  \begin{quantikz} 
    \ket{0^{3}}  & \gate[1]{H^{\otimes 3}}{} &\ctrl{1} & \qw &\qw &\qw \\
    \ket{0^{3}}  & \qw & \targ{} & \qw &\qw &\qw \\
\end{quantikz}
=\begin{quantikz} 
    \ket{0}  & \gate[1]{H}{} &\ctrl{3} & \qw &\qw &\qw \\
        \ket{0}  & \gate[1]{H}{} &\qw & \ctrl{3} &\qw &\qw \\
            \ket{0}  & \gate[1]{H}{} &\qw & \qw &\ctrl{3} &\qw \\
    \ket{0}  & \qw & \targ{} & \qw &\qw &\qw \\
        \ket{0}  & \qw & \qw & \targ{} &\qw &\qw \\
            \ket{0}  & \qw & \qw & \qw &\targ{} & \qw \\
\end{quantikz}
\caption{Illustration of the multi-qubit CNOT gate used to produce $\ket{\chi}$ in Figure \ref{fig:doscircuit} with $n=3$.}
\label{fig:multiCNOT}
\end{figure}
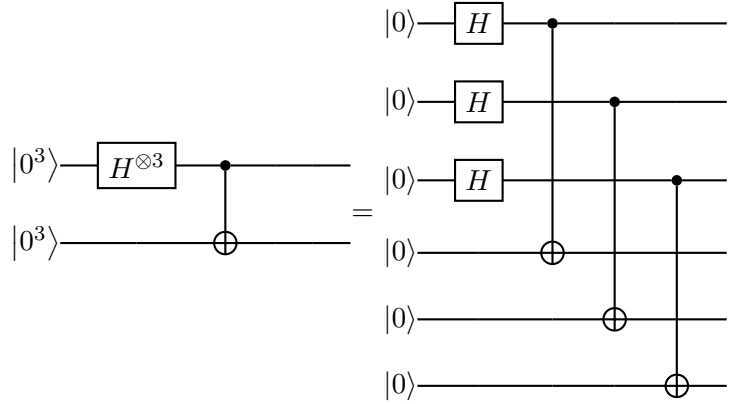

The first two layers of the circuit shown in Figure~\ref{fig:doscircuit} are used to prepare $\ket{\chi} = \frac{1}{\sqrt{N}}\sum_{x} \ket{x}\ket{x}$ needed to produce $\ket{\psi_i}=\frac{1}{\sqrt{N}}\sum_{x=0}^{N-1} \sqrt{g( \mathcal{H}'_{\rm pair} -\omega_i I)} \ket{x}\ket{x}$. In this part of the circuit, we first apply $n$ Hadamard gates to the first set of $n$ qubits initialized as $\ket{0}$ to obtain the state $\frac{1}{\sqrt{N}}\sum_{x} \ket{x}\ket{0^n}$. We then use a series of CNOT gates placed on matching qubits to bring the second set of $n$ qubits to be in the superposition of $\ket{x}$ also.  Instead of measuring several ancilla qubits in Figure~\ref{fig:doscircuit}, additional CNOT gates can be used to reduce the number of ancilla qubits to be measured to 1. 
While the focus of this paper is on the circuit design for block encoding the second-quantized pairing Hamiltonian, we hold the discussions of the initial state preparation and measurement to future works.

\begin{figure}
\centering
   \includegraphics[width=0.7\linewidth]{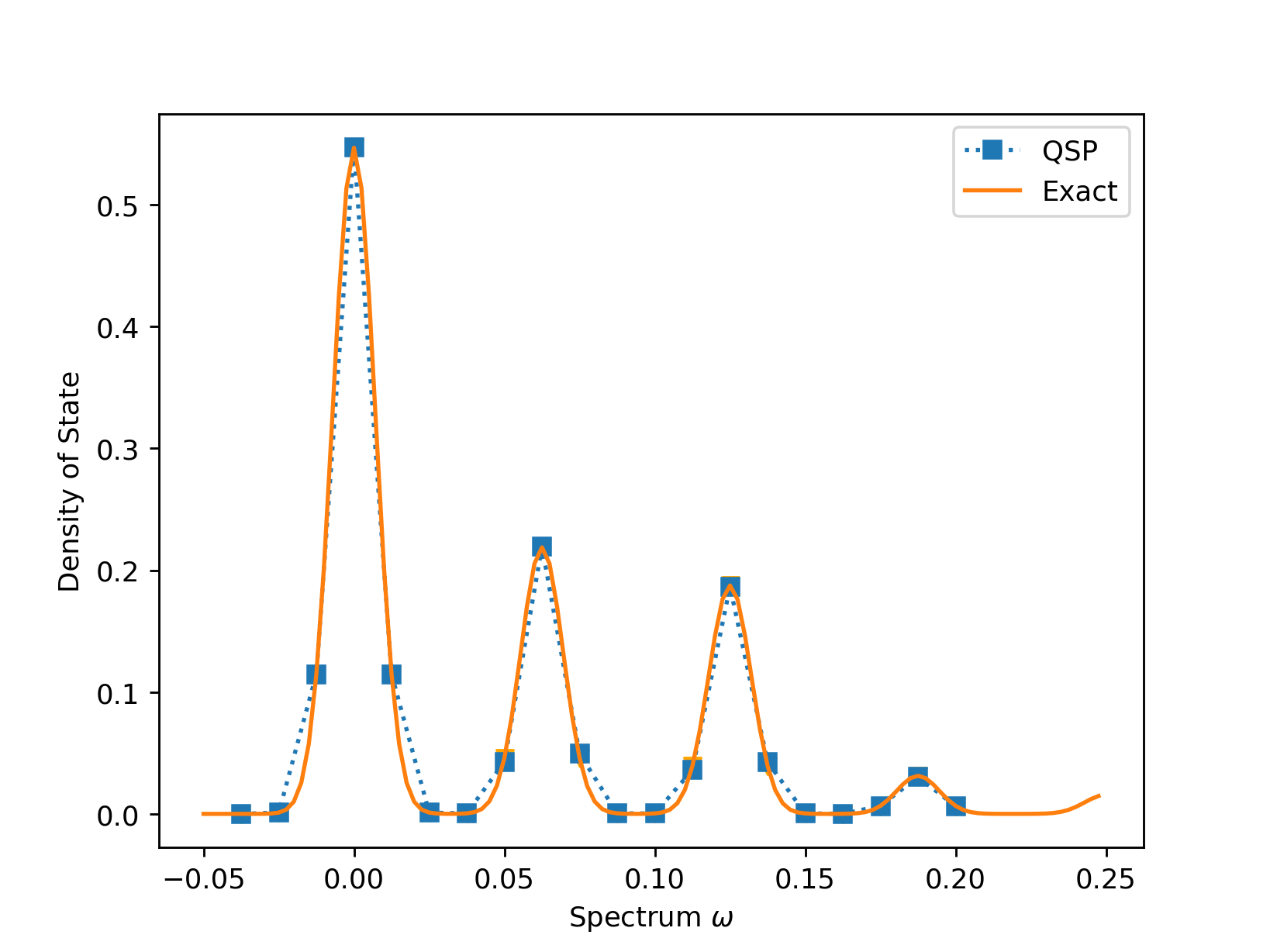}
\caption{The density of states of the pairing Hamiltonian for all the possible many-nucleon systems supported by the restricted basis set.
The solid line indicates the exact density of states via classical calculations.
The squares connected by the dashed lines indicate the density of states from quantum computation.
We employ $(d_{qsp}+1)$-th
degree polynomial to approximate $\tilde{g}(\lambda-\omega_j)$. The QSP calculation is performed with $d_{qsp}=50$ for good convergence.} 
\label{fig:dos}
\end{figure}

Figure \ref{fig:dos} shows the comparison between DOS obtained from the diagonalization of $\sH '_{\rm pair}$ and evaluating \eqref{eq:tracedos}, which is shown as the solid curve, and the square amplitudes of $\psi_i$'s defined by \eqref{eq:psinorm2} in which a polynomial of degree $d_{qsp}=50$ is constructed (via QSP) to approximate $\sqrt{ {\tilde g}(\lambda-\omega_i)}$, which are shown as solid square markers connected by a dotted curve.  Because the circuit we construct produces a $(16,5)$-block encoding of $\sH_{\rm pair}$, the computed DOS is the DOS of $\sH_{\rm pair}$/16.  We use the same $\tilde g$ in both calculations. The former can be viewed as an exact DOS calculation and the latter can be interpreted as the DOS approximation obtained from the quantum algorithm presented in this section. 
This figure shows that the approximate DOS obtained from the quantum algorithm match well with exact DOS. All major peaks of the DOS can be correctly identified from the dotted curve within the spectrum interval of interest $[-0.05,0.2]$.


As we indicated in section~\ref{sec:circuit}, the gate complexity for constructing a block encoding quantum circuit $U_B$ for $ \sH '_{\rm pair} $ is $\mathcal{O}(\text{poly}(n))$.  The number of $U_B$ blocks and the number of additional CNOT and phase gates in the QSVT circuit shown in Figure~\ref{fig:QSP_example} depends on the degree of the polynomial required to accurately approximate $\tilde{g}(\omega)$, which in turn depends on the spectrum width and the required resolution of the DOS. With the degree of polynomial $d_{qsp}$ in QSP being $\mathcal{O}(\text{poly}(n))$, the overall gate complexity for the quantum circuit associated with each $\omega_i$ is $\mathcal{O}(\text{poly}(n))$.

The query complexity of the quantum DOS approximation depends on
the number of energy samples $\omega_i$'s required to construct the DOS approximation and the number of measurements necessary for sufficient success measurements. If both scale as $\mathcal{O}(\text{poly}(n))$, then the overall complexity of the quantum DOS approximation is $\mathcal{O}(\text{poly}(n))$, which is superior to the $\mathcal{O}(\text{poly}(N))$ required in a classical DOS estimation algorithm~\cite{lin2016approximating}.  However, for the restricted problem presented in this example, the total number of gates required in the circuit shown in Figure~\ref{fig:doscircuit} exceeds $N=64$ due to restricted problem size that is far from the asymptotic regime. A more detailed analysis of the quantum resources required to obtain a quantum DOS approximation is beyond the scope of this paper and will be addressed in future researches.

\section{Conclusion}
\label{sec:conclude}
In this paper, we follow the general strategy presented in~\cite{camps2022explicit} to develop a direct and explicit way to construct a block encoding circuit for a nuclear pairing Hamiltonian by treating it as a sparse matrix with a special structure. The circuit consists of an $O_C$ circuit block that encodes the sparsity structure of the Hamiltonian and an $O_\sH$ circuit block that encodes the numerical values of the nonzero matrix elements in the Hamiltonian.  The $O_C$ block contains a sequence of controlled swap operations combined with controlled validation operations.  The $O_\sH$ block contains a sequence of controlled rotations.  The gate complexity of the proposed block encoding circuit is $\mathcal{O}(\text{poly}(n))$ with $n$ being the number of single-particle states used to represent the many-body wavefunctions associated with the Hamiltonian. The number of ancilla qubits required is $\log{L}+3$ where $L$ is the number of terms in \eqref{eq:hpairnumber}. In practice, $L$ is typically a polynomial in $n$. Hence the number of ancilla qubits is $\mathcal{O}(\log(n))$.  
The gate complexity and ancilla qubit count are comparable to that required in a circuit that block encodes the pairing Hamiltonian after it is written as a linear combination of (unitary) Pauli operators (strings). We refer to this approach as the LCU block encoding scheme.
We believe that being able to directly construct a block encoding circuit without resorting to mapping the creation and annihilation operators to the Pauli strings through the Jordan-Wigner (JW) or Bravyi-Kitaev (BK) transformations simplifies the circuit construction process and is more intuitive and clear. 
Utilizing the JW or BK transformation creates $16$ Pauli strings for each two-body interaction term in the pairing Hamiltonian \eqref{eq:hpairnumber}. One additional qubit is needed to account for the linear combination of these Pauli strings.  

We have not performed a thorough analysis of the query complexity and success probability of measurement in the paper. However, we expect the success probability of measuring $\ket{0}$ from the output of the circuit presented here to be comparable to that associated with the block encoding circuit for the linear combination of Pauli strings since the $l_1$ norm of coefficients in both approaches are the same. 

Although the $O_C$ circuit block that consists of a sequence of controlled swap operations is comparable in structure to the select oracle used in the LCU based block encoding circuit,  the $O_\sH$ circuit block that contains a sequence of controlled rotations is considerably easier to construct in general than the prepare oracle circuit required in an LCU based approach.
The prepare oracle in the LCU based approach requires encoding $8L$ coefficients that appear in the method presented in this paper.  The advantage of constructing a circuit for $O_{\sH}$ over a prepare oracle in LCU is also expected for a more general $k$-body Hamiltonian for $k>2$.

%

We also note that, even though we have demonstrated a direct and explicit block encoding circuit construction for a second-quantized pairing Hamiltonian, the technique presented here is rather general and can be extended to any second-quantized many-body Hamiltonian.  For a more general second-quantized Hamiltonian, the control used for the swap operation can potentially be more complex. However, there should not be any fundamental difficulties in implementing such a circuit block. The main issue that needs to be properly addressed is the presence of additional phase terms in the $c(j,l)$ function. These phase terms can be encoded by utilizing previously developed techniques that can be found in, for example, ~\cite{wan2021exponentially}. We also refer readers to works~\cite{du2023multi,du2024hamiltonian} on addressing the phase terms in general second-quantized Hamiltonians of relativistic and nonrelativistic many-fermion systems.

\appendix

\section*{Acknowledgments}
 This work was supported by US DOE Grant DE-SC0023707 under the Office of Nuclear Physics Quantum Horizons program for the "{\bf Nu}Nuclei and {\bf Ha}drons with {\bf Q}uantum computers ({\bf NuHaQ})" project (W.D., J.V.). This work was supported by Computing Sciences 2023 Summer Program at Berkeley Lab (D.L.). This work was also supported by the Quantum Systems Accelerator program  (L.L., C.Y.),  by  the  Applied Mathematics Program of the US Department of Energy (DOE) Office of Advanced Scientific Computing Research under contract number DE-AC02-05CH1123 (L.L.), the DOE Office of Advanced Scientific Computing Research Accelerated Research for Quantum Computing (ARQC) Program (C.Y.) under U.S. Department of Energy Contract No. DE-AC02-05CH11231. L.L. is a Simons Investigator in Mathematics.

\bibdata{bib}
\bibliographystyle{unsrt}
\bibliography{bio}

\begin{thebibliography}{10}

\bibitem{camps2022explicit}
Daan Camps, Lin Lin, Roel Van~Beeumen, and Chao Yang.
\newblock Explicit quantum circuits for block encodings of certain sparse
  matrices.
\newblock {\em arXiv preprint arXiv:2203.10236}, 2022.

\bibitem{low2017optimal}
Guang~Hao Low and Isaac~L Chuang.
\newblock Optimal hamiltonian simulation by quantum signal processing.
\newblock {\em Physical review letters}, 118(1):010501, 2017.

\bibitem{low2019hamiltonian}
Guang~Hao Low and Isaac~L Chuang.
\newblock Hamiltonian simulation by qubitization.
\newblock {\em Quantum}, 3:163, 2019.

\bibitem{gilyen2019quantum}
Andr{\'a}s Gily{\'e}n, Yuan Su, Guang~Hao Low, and Nathan Wiebe.
\newblock Quantum singular value transformation and beyond: exponential
  improvements for quantum matrix arithmetics.
\newblock In {\em Proceedings of the 51st Annual ACM SIGACT Symposium on Theory
  of Computing}, pages 193--204, 2019.

\bibitem{nielsen2001quantum}
Michael~A Nielsen and Isaac~L Chuang.
\newblock Quantum computation and quantum information.
\newblock {\em Phys. Today}, 54(2):60, 2001.

\bibitem{lin2022lecture}
Lin Lin.
\newblock Lecture notes on quantum algorithms for scientific computation.
\newblock {\em arXiv preprint arXiv:2201.08309}, 2022.

\bibitem{childs2017lecture}
Andrew~M Childs.
\newblock Lecture notes on quantum algorithms.
\newblock {\em Lecture notes at University of Maryland}, 2017.

\bibitem{sunderhauf2023block}
Christoph S{\"u}nderhauf, Earl Campbell, and Joan Camps.
\newblock Block-encoding structured matrices for data input in quantum
  computing.
\newblock {\em arXiv preprint arXiv:2302.10949}, 2023.

\bibitem{WanBlock}
Lin-Chun Wan, Chao-Hua Yu, Shi-Jie Pan, Su-Juan Qin, Fei Gao, and Qiao-Yan Wen.
\newblock Block-encoding-based quantum algorithm for linear systems with
  displacement structures.
\newblock {\em Phys. Rev. A}, 104:062414, Dec 2021.

\bibitem{loke2017efficient}
Thomas Loke and Jingbo~B Wang.
\newblock Efficient quantum circuits for szegedy quantum walks.
\newblock {\em Annals of Physics}, 382:64--84, 2017.

\bibitem{camps2022algebraic}
Daan Camps, Efekan Kökcü, Lindsay Bassman~Oftelie, Wibe~A De~Jong,
  Alexander~F Kemper, and Roel Van~Beeumen.
\newblock An algebraic quantum circuit compression algorithm for hamiltonian
  simulation.
\newblock {\em SIAM Journal on Matrix Analysis and Applications},
  43(3):1084--1108, 2022.

\bibitem{camps2022fable}
Daan Camps and Roel Van~Beeumen.
\newblock Fable: Fast approximate quantum circuits for block-encodings.
\newblock In {\em 2022 IEEE International Conference on Quantum Computing and
  Engineering (QCE)}, pages 104--113. IEEE, 2022.

\bibitem{camps2020approximate}
Daan Camps and Roel Van~Beeumen.
\newblock Approximate quantum circuit synthesis using block encodings.
\newblock {\em Physical Review A}, 102(5):052411, 2020.

\bibitem{leng2024expanding}
Jiaqi Leng, Joseph Li, Yuxiang Peng, and Xiaodi Wu.
\newblock Expanding hardware-efficiently manipulable hilbert space via
  hamiltonian embedding.
\newblock {\em arXiv preprint arXiv:2401.08550}, 2024.

\bibitem{zhou2017efficient}
SS~Zhou and JB~Wang.
\newblock Efficient quantum circuits for dense circulant and circulant like
  operators.
\newblock {\em Royal Society open science}, 4(5):160906, 2017.

\bibitem{li2023efficient}
Haoya Li, Hongkang Ni, and Lexing Ying.
\newblock On efficient quantum block encoding of pseudo-differential operators.
\newblock {\em Quantum}, 7:1031, 2023.

\bibitem{wan2021exponentially}
Kianna Wan.
\newblock Exponentially faster implementations of select (h) for fermionic
  hamiltonians.
\newblock {\em Quantum}, 5:380, 2021.

\bibitem{babbush2018encoding}
Ryan Babbush, Craig Gidney, Dominic~W Berry, Nathan Wiebe, Jarrod McClean,
  Alexandru Paler, Austin Fowler, and Hartmut Neven.
\newblock Encoding electronic spectra in quantum circuits with linear t
  complexity.
\newblock {\em Physical Review X}, 8(4):041015, 2018.

\bibitem{du2023multi}
Weijie Du and James~P. Vary.
\newblock {Multinucleon structure and dynamics via quantum computing}.
\newblock {\em Phys. Rev. A}, 108(5):052614, 2023.

\bibitem{babbush2017low}
Ryan Babbush, Nathan Wiebe, Jarrod McClean, James McClain, Hartmut Neven, and
  Garnet~Kin Chan.
\newblock Low depth quantum simulation of electronic structure.
\newblock {\em arXiv preprint arXiv:1706.00023}, 2017.

\bibitem{babbush2018low}
Ryan Babbush, Nathan Wiebe, Jarrod McClean, James McClain, Hartmut Neven, and
  Garnet Kin-Lic Chan.
\newblock Low-depth quantum simulation of materials.
\newblock {\em Physical Review X}, 8(1):011044, 2018.

\bibitem{chan2023simulating}
Hans Hon~Sang Chan, David Mu{\~n}oz-Ramo, and Nathan Fitzpatrick.
\newblock Simulating non-unitary dynamics using quantum signal processing with
  unitary block encoding.
\newblock {\em arXiv preprint arXiv:2303.06161}, 2023.

\bibitem{nielsen2005fermionic}
Michael~A Nielsen et~al.
\newblock The fermionic canonical commutation relations and the jordan-wigner
  transform.
\newblock {\em School of Physical Sciences The University of Queensland}, 59,
  2005.

\bibitem{seeley2012bravyi}
Jacob~T Seeley, Martin~J Richard, and Peter~J Love.
\newblock The bravyi-kitaev transformation for quantum computation of
  electronic structure.
\newblock {\em The Journal of chemical physics}, 137(22), 2012.

\bibitem{tranter2018comparison}
Andrew Tranter, Peter~J Love, Florian Mintert, and Peter~V Coveney.
\newblock A comparison of the bravyi--kitaev and jordan--wigner transformations
  for the quantum simulation of quantum chemistry.
\newblock {\em Journal of chemical theory and computation}, 14(11):5617--5630,
  2018.

\bibitem{Jensen2017}
Maria Paola~Lombardo Morten Hjorth-Jensen and Ubirajara van Kolck.
\newblock {\em An Advanced Course in Computational Nuclear Physics: Bridging
  the Scales from Quarks to Neutron Stars (Lecture Notes in Physics, 936)}.
\newblock Springer, 2017.
\newblock 1st ed.

\bibitem{Suhonen2010}
Jouni Suhonen.
\newblock {\em From Nucleons to Nucleus: Concepts of Microscopic Nuclear
  Theory}.
\newblock Springer, 2010.
\newblock 1st ed.

\bibitem{GrandUni}
John~M. Martyn, Zane~M. Rossi, Andrew~K. Tan, and Isaac~L. Chuang.
\newblock Grand unification of quantum algorithms.
\newblock {\em PRX Quantum}, 2:040203, Dec 2021.

\bibitem{nielsen2010quantum}
M.A. Nielsen and I.L. Chuang.
\newblock {\em Quantum Computation and Quantum Information: 10th Anniversary
  Edition}.
\newblock Cambridge University Press, 2010.

\bibitem{kay2018tutorial}
Alastair Kay.
\newblock Tutorial on the quantikz package.
\newblock {\em arXiv preprint arXiv:1809.03842}, 2018.

\bibitem{dong2022infinite}
Yulong Dong, Lin Lin, Hongkang Ni, and Jiasu Wang.
\newblock Infinite quantum signal processing.
\newblock {\em arXiv preprint arXiv:2209.10162}, 2022.

\bibitem{Dongefficient}
Yulong Dong, Xiang Meng, K.~Birgitta Whaley, and Lin Lin.
\newblock Efficient phase-factor evaluation in quantum signal processing.
\newblock {\em Phys. Rev. A}, 103:042419, Apr 2021.

\bibitem{vale2023decomposition}
Rafaella Vale, Thiago Melo~D Azevedo, Ismael Ara{\'u}jo, Israel~F Araujo, and
  Adenilton~J da~Silva.
\newblock Decomposition of multi-controlled special unitary single-qubit gates.
\newblock {\em arXiv preprint arXiv:2302.06377}, 2023.

\bibitem{shende2008cnot}
Vivek~V Shende and Igor~L Markov.
\newblock On the cnot-cost of toffoli gates.
\newblock {\em arXiv preprint arXiv:0803.2316}, 2008.

\bibitem{Du:2023bpw}
Weijie Du and James~P. Vary.
\newblock {Multinucleon structure and dynamics via quantum computing}.
\newblock {\em Phys. Rev. A}, 108(5):052614, 2023.

\bibitem{lin2016approximating}
Lin Lin, Yousef Saad, and Chao Yang.
\newblock Approximating spectral densities of large matrices.
\newblock {\em SIAM review}, 58(1):34--65, 2016.

\bibitem{Wang2022energylandscapeof}
Jiasu Wang, Yulong Dong, and Lin Lin.
\newblock On the energy landscape of symmetric quantum signal processing.
\newblock {\em {Quantum}}, 6:850, November 2022.

\bibitem{du2024hamiltonian}
Weijie Du and James~P Vary.
\newblock Hamiltonian input model and spectroscopy on quantum computers.
\newblock {\em arXiv preprint arXiv:2402.08969}, 2024.

\end{thebibliography}

\end{document}